\documentclass[aps,prd,superscriptaddress,preprint,tightenlines,nofootinbib,floatfix]{revtex4}

\usepackage{graphicx} 
\usepackage{dcolumn}  
\usepackage{bm}       

\def \bea{\begin{eqnarray}}
\def \beq{\begin{equation}}

\def \eea{\end{eqnarray}}
\def \eeq{\end{equation}}

\newcommand{\gev}{ \ensuremath{\,\mathrm{GeV}}}
\newcommand{\mev}{ \ensuremath{\,\mathrm{MeV}}}

\newcommand{\eg}{\textit{e.g.}}

\newcommand{\psipr}{{\ensuremath{\psi'}}}
\newcommand{\jpsi}{{\ensuremath{J/\psi}}}

\newcommand{\costhp}{\ensuremath{\cos \theta'}}
\newcommand{\costh}{\ensuremath{\cos \theta}}

\newcommand{\costhgg}{\ensuremath{\cos \theta_{\gamma \gamma'}}}

\renewcommand{\vec}[1]{\ensuremath{{\mathbf{#1}}}}
\newcommand{\pow}[1]{\ensuremath{\times 10^{#1}}}  
  
\newcommand{\tsb}[1]{\ensuremath{_{\mathrm{#1}}}}  
\newcommand{\tsp}[1]{\ensuremath{^{\mathrm{#1}}}}  
\newcommand{\phip}{\ensuremath{\phi'}}

\newcommand{\chic}[1]{\ensuremath{\chi_{c #1}}}

\newcommand{\psip}{\ensuremath{\psi'}}
\newcommand{\gamp}{\ensuremath{\gamma'}}
\newcommand{\gamu}{\ensuremath{\gamma}}

\newcommand{\ket}[1]{\ensuremath{\left| #1 \right \rangle}}

\newcommand{\pmbA}[0]{{\cal A}}
\newcommand{\PDF}[0]{\ensuremath{W(\Omega; \pmbA)}}
\newcommand{\PDFz}[0]{\ensuremath{W(\Omega; \pmbA_0)}}
\newcommand{\fivevars}{\ensuremath{\costhp, \phip, \costhgg, \costh, \phi}}

\begin{document}

\preprint{CLNS 09/2059}       
\preprint{CLEO 09-12~~~}         

\title{Higher-order multipole amplitudes in charmonium radiative transitions}

\author{M.~Artuso}
\author{S.~Blusk}
\author{S.~Khalil}
\author{R.~Mountain}
\author{K.~Randrianarivony}
\author{T.~Skwarnicki}
\author{S.~Stone}
\author{J.~C.~Wang}
\author{L.~M.~Zhang}
\affiliation{Syracuse University, Syracuse, New York 13244, USA}
\author{G.~Bonvicini}
\author{D.~Cinabro}
\author{A.~Lincoln}
\author{M.~J.~Smith}
\author{P.~Zhou}
\author{J.~Zhu}
\affiliation{Wayne State University, Detroit, Michigan 48202, USA}
\author{P.~Naik}
\author{J.~Rademacker}
\affiliation{University of Bristol, Bristol BS8 1TL, UK}
\author{D.~M.~Asner}
\author{K.~W.~Edwards}
\author{J.~Reed}
\author{A.~N.~Robichaud}
\author{G.~Tatishvili}
\author{E.~J.~White}
\affiliation{Carleton University, Ottawa, Ontario, Canada K1S 5B6}
\author{R.~A.~Briere}
\author{H.~Vogel}
\affiliation{Carnegie Mellon University, Pittsburgh, Pennsylvania 15213, USA}
\author{P.~U.~E.~Onyisi}
\author{J.~L.~Rosner}
\affiliation{University of Chicago, Chicago, Illinois 60637, USA}
\author{J.~P.~Alexander}
\author{D.~G.~Cassel}
\author{R.~Ehrlich}
\author{L.~Fields}
\author{R.~S.~Galik}
\author{L.~Gibbons}
\author{S.~W.~Gray}
\author{D.~L.~Hartill}
\author{B.~K.~Heltsley}
\author{J.~M.~Hunt}
\author{D.~L.~Kreinick}
\author{V.~E.~Kuznetsov}
\author{J.~Ledoux}
\author{H.~Mahlke-Kr\"uger}
\author{J.~R.~Patterson}
\author{D.~Peterson}
\author{D.~Riley}
\author{A.~Ryd}
\author{A.~J.~Sadoff}
\author{X.~Shi}
\author{S.~Stroiney}
\author{W.~M.~Sun}
\affiliation{Cornell University, Ithaca, New York 14853, USA}
\author{J.~Yelton}
\affiliation{University of Florida, Gainesville, Florida 32611, USA}
\author{P.~Rubin}
\affiliation{George Mason University, Fairfax, Virginia 22030, USA}
\author{N.~Lowrey}
\author{S.~Mehrabyan}
\author{M.~Selen}
\author{J.~Wiss}
\affiliation{University of Illinois, Urbana-Champaign, Illinois 61801, USA}
\author{M.~Kornicer}
\author{R.~E.~Mitchell}
\author{M.~R.~Shepherd}
\author{C.~M.~Tarbert}
\affiliation{Indiana University, Bloomington, Indiana 47405, USA }
\author{D.~Besson}
\affiliation{University of Kansas, Lawrence, Kansas 66045, USA}
\author{T.~K.~Pedlar}
\author{J.~Xavier}
\affiliation{Luther College, Decorah, Iowa 52101, USA}
\author{D.~Cronin-Hennessy}
\author{K.~Y.~Gao}
\author{J.~Hietala}
\author{R.~Poling}
\author{P.~Zweber}
\affiliation{University of Minnesota, Minneapolis, Minnesota 55455, USA}
\author{S.~Dobbs}
\author{Z.~Metreveli}
\author{K.~K.~Seth}
\author{B.~J.~Y.~Tan}
\author{A.~Tomaradze}
\affiliation{Northwestern University, Evanston, Illinois 60208, USA}
\author{S.~Brisbane}
\author{J.~Libby}
\author{L.~Martin}
\author{A.~Powell}
\author{P.~Spradlin}
\author{C.~Thomas}
\author{G.~Wilkinson}
\affiliation{University of Oxford, Oxford OX1 3RH, UK}
\author{H.~Mendez}
\affiliation{University of Puerto Rico, Mayaguez, Puerto Rico 00681}
\author{J.~Y.~Ge}
\author{D.~H.~Miller}
\author{I.~P.~J.~Shipsey}
\author{B.~Xin}
\affiliation{Purdue University, West Lafayette, Indiana 47907, USA}
\author{G.~S.~Adams}
\author{D.~Hu}
\author{B.~Moziak}
\author{J.~Napolitano}
\affiliation{Rensselaer Polytechnic Institute, Troy, New York 12180, USA}
\author{K.~M.~Ecklund}
\affiliation{Rice University, Houston, Texas 77005, USA}
\author{J.~Insler}
\author{H.~Muramatsu}
\author{C.~S.~Park}
\author{E.~H.~Thorndike}
\author{F.~Yang}
\affiliation{University of Rochester, Rochester, New York 14627, USA}
\collaboration{CLEO Collaboration}
\noaffiliation


\date{October 1, 2009}

\begin{abstract}

Using 24 million $\psipr \equiv \psi(2S)$ decays in CLEO-c, we have searched
for higher
multipole admixtures in electric-dipole-dominated radiative transitions in
charmonia.  We find good agreement between our data and theoretical predictions
for magnetic quadrupole ($M2$) amplitudes in the transitions $\psi' \to \gamma
\chi_{c1,c2}$ and $\chi_{c1,c2} \to \gamma J/\psi$, in striking contrast to
some previous measurements.  Let $b_2^J$ and $a_2^J$ denote the normalized $M2$
amplitudes in the respective aforementioned decays, where
the superscript $J$ refers to the angular momentum of the $\chi_{cJ}$.  By
performing unbinned maximum likelihood fits to full five-parameter angular
distributions, we found the following values of $M2$ admixtures for
 $J_\chi\!=\!1$:
  $a_2^{J = 1} = (-6.26 \pm 0.63 \pm 0.24)\pow{-2}$ and
  $b_2^{J = 1} = (2.76 \pm 0.73 \pm 0.23)\pow{-2}$, 
which agree well with theoretical expectations for a vanishing anomalous
magnetic moment of the charm quark. For $J_\chi \!=\!2$, if we fix the electric
octupole ($E3$) amplitudes to zero as theory predicts for transitions between
charmonium S states and P states, we find 
$a_2^{J=2} = (-9.3 \pm 1.6 \pm 0.3)\pow{-2}$ and 
$b_2^{J=2} = (1.0 \pm 1.3 \pm 0.3) \pow{-2}$.  
If we allow for $E3$ amplitudes we find, with a four-parameter fit,
$a_2^{J=2} = (-7.9 \pm 1.9 \pm 0.3)\pow{-2}$,
  $b_2^{J=2}= (0.2 \pm 1.4 \pm 0.4)\pow{-2}$,
  $a_3^{J=2}= (1.7 \pm 1.4 \pm 0.3)\pow{-2}$, and
  $b_3^{J=2} =(-0.8 \pm 1.2 \pm 0.2)\pow{-2}$.
We determine the ratios $a_2^{J=1}/a_2^{J=2} = 0.67^{+0.19}_{-0.13}$ and
$a_2^{J=1}/b_2^{J=1} = -2.27^{+0.57}_{-0.99}$, where the theoretical
predictions are independent of the charmed quark magnetic moment and are
$a_2^{J=1}/a_2^{J=2} = 0.676 \pm 0.071$ and $a_2^{J=1}/b_2^{J=1} = -2.27
\pm 0.16$.
\end{abstract}

\maketitle

\section{INTRODUCTION \label{sec:intro}}

\begin{figure}[htb]
\begin{center}
\includegraphics[width=4in]{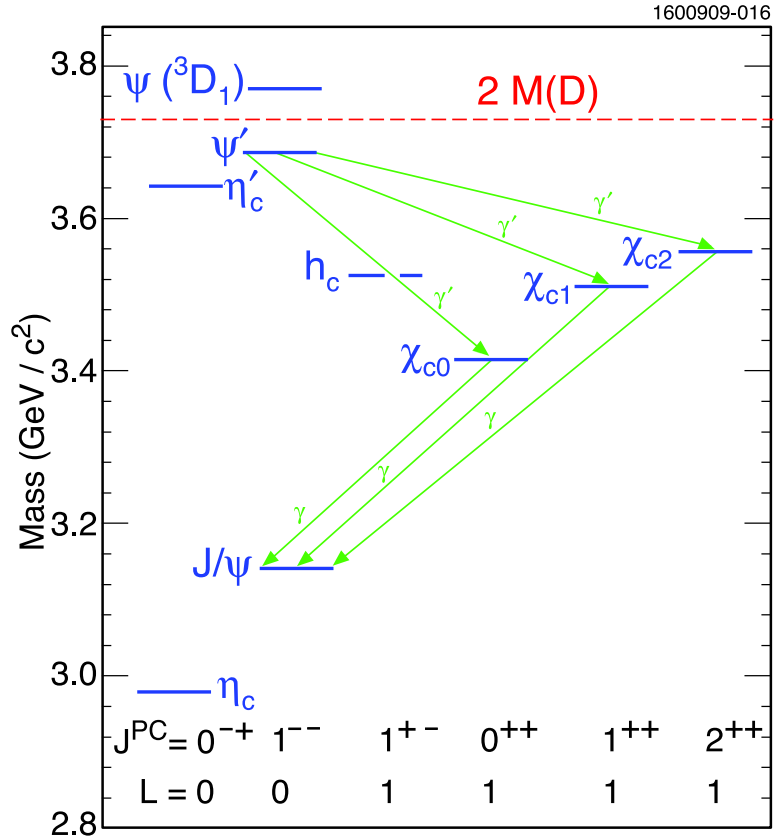}
\end{center}
\caption{Charmonium energy levels. Only the transitions studied in this 
article are shown.
\label{fig:energy_level_diagram}}
\end{figure}

The radiative transitions between spin-triplet charmonium states are known to be
dominated by electric dipole ($E1$) amplitudes, but higher multipole
contributions, magnetic quadrupole and electric octupole 
($M2$ and $E3$), are sometimes allowed.  These higher multipoles give
information about the magnetic moment of the charm quark.  To search for these
contributions, we studied the radiative decay sequences
\bea
 e^+ e^- \to \gamma^* &\to & \psip \equiv \psi(2S) \nonumber \\
 \psip & \to & \gamp \chi_{(c1,c2)} \nonumber \\
 \chi_{(c1,c2)} & \to & \gamu \jpsi \nonumber \\
 \jpsi & \to & e^+ e^- ~{\rm or}~ \mu^+ \mu^- \nonumber 
\eea
using the helicity formalism developed in Refs.\
\cite{KMR:1976,KMR:1980,Rosner:2008,BC:1976}.  As shown in Fig.\
\ref{fig:energy_level_diagram}, the particles $\psip$, $\chi_{(c1,c2)}$, and
$\jpsi$ are the $2\, ^{3}S_1$, $\;1 ^{3}P_{(1,2)}$, and $1\, ^{3}S_1$
charmonium states, respectively.  For the $J_\chi = 1$ decay sequence, we
search for two multipole amplitudes, $b_2^{J=1}$ and $a_2^{J=1}$, which
are respectively the $M2$ amplitudes for the $\psip \to \gamp \chi_{c1}$
($b$ for $b$efore the $\chi_c$) and $\chi_{c1} \to \gamu \jpsi$ decay
($a$ for $a$fter the $\chi_c$).  Similarly, for the $J_\chi
= 2$ decay sequence, we search for two $M2$ amplitudes ($b_2^{J=2}$, $a_2^{J=2}$)
and two $E3$ amplitudes ($b_3^{J=2}$, $a_3^{J=2}$).


\begin{figure}[htb]
\includegraphics[width=4.8in]{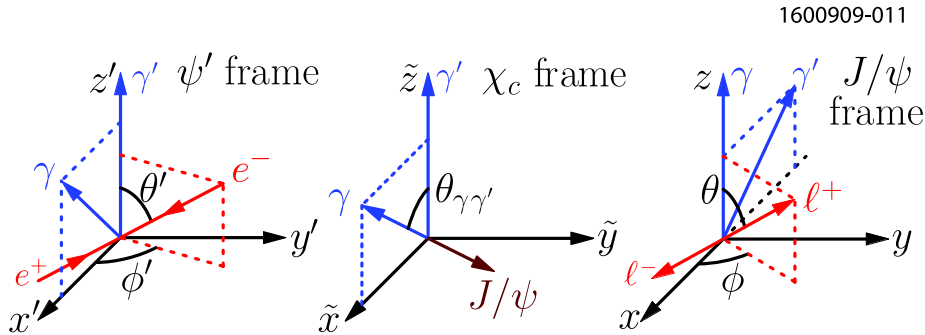}
\caption{Reference frames defining the angles used in this analysis.  In the
$\psi'$ frame, the angles $\theta'$, $\phi'$ are the polar and
azimuthal angles of the beam pipe (specifically, the positron's direction)
relative to $\gamma'$ defining the $z'$-axis, and $\gamma$ lying in the $x'$-$z'$
plane (with a \textit{positive} $x'$-component). In the $\chi_c$ frame, the
angle $\theta_{\gamma \gamma'}$ is the angle between the two photons. In the
$J / \psi$ frame, the angles $\theta$, $\phi$ are the polar and
azimuthal angles of the two leptons (specifically, the positive lepton's
direction) relative to $\gamma$ defining the $z$-axis, and $\gamma'$ lying in
the $x$-$z$ plane (with a \textit{negative} $x$-component).}
\label{fig:angles}
\end{figure}

The multipole amplitudes are calculated from a maximum likelihood fit of the
joint angular distribution of the two photons $\gamma'$ and $\gamma$, described
by five angles for each event.  Polar and azimuthal angles $(\theta',\phi')$
denote the direction of the initial $e^+ e^-$ axis relative to the first
photon $\gamma'$ (in the $\psipr$ frame), an angle $\theta_{\gamma \gamma'}$
describes the direction between the two photons (in the $\chi_c$ frame), and
polar and azimuthal angles $(\theta,\phi)$ denote the direction of the final
lepton pair
($\ell^+ \ell^-$) axis relative to the second photon $\gamma$ (in the $\jpsi$
frame).  These angles are illustrated in Fig.~\ref{fig:angles}.


In previous experimental studies of $\chi_{cJ} \to \gamma \jpsi$, the magnetic
quadrupole amplitude in the decay sequences involving $\chi_{c1}$ was found
to be consistent with zero, while that found via $\chi_{c2}$ was found to be
several standard deviations from zero.  However, theory predicts the ratio of
these two magnetic quadrupole amplitudes to be of order unity.  With
CLEO's large sample of $\psi'$ decays, the question is ripe for
re-investigation.  The present paper describes that effort.

Section \ref{sec:theor} sets the theoretical stage for the investigation.
Prior experimental results are reviewed in Sec.\ \ref{sec:prior}.  The CLEO
detector, data sets, and Monte Carlo samples are described in Sec.\
\ref{sec:detdtmc}.  Sec.\ \ref{selection_criteria} discusses selection
criteria, while Sec.\ \ref{sec:fits} is
devoted to fits to the data.  Systematic uncertainties are treated in
Sec.\ \ref{sec:sys}, while Sec.\ \ref{sec:concl} concludes.

\section{THEORETICAL CONTEXT \label{sec:theor}}

\subsection{Allowed radiative transitions}

For the radiative decays between a $^3 S_1$ state and a $^3 P_{1}$ state, only
$E1$ and $M2$ transitions are allowed. For ${^3S_1} \to {^3P_2}$ transitions,
from conservation of angular momentum and parity, we would expect that the $E3$ 
transition would be allowed, but this is forbidden under the single-quark  
radiation hypothesis \cite{KMR:1980,OSMS:1985}.  Single-quark radiative 
transitions must have $|\Delta S| \leq 1$ and parity-changing transitions must
have $|\Delta L| = 1$, so the photon cannot carry off three units of angular
momentum \cite{KMR:1976}.  However, for the $J_\chi = 2$ case, electric
octupole transitions are allowed if either $S$ state has a small $D$ admixture
\cite{QWG}, or if the $P$ state has a small $F$ admixture.
There is evidence \cite{QWG,Rosner:2001,Rosner:2004wy} that the $\psip$ state
is actually a mixture $\cos \varphi \ket{2\; ^3 S_1} - \sin \varphi
\ket{^3 D_1}$ with $\varphi = (12 \pm 2)^\circ$, so we may expect a small $b_3$
transition amplitude.

\subsection{Joint angular distribution}

The formalism developed in Refs.\ \cite{KMR:1976, KMR:1980, Richman:1984} is
used to construct the joint angular distribution of the decay sequence.  We
denote the signal decay as 
\bea
  \psipr (\lambda') & \to & \gamma' (\mu') + \chi  (\nu') \label{eq:sigB} \\
   \chi (\nu) & \to & \gamma (\mu) + \jpsi (\lambda)  \label{eq:sigA}
\eea
with helicities in parentheses and the helicities associated with the $\psipr$
decay labeled with primes.  For the $\psipr$ ($\chi_c$) decay sequence, the
helicity amplitudes are labeled $B_{\nu'}$ ($A_{\nu}$) and the multipole
amplitudes are labeled $b_{J_\gamma}$ ($a_{J_{\gamma'}}$), where $J_\gamma$ is
the angular momentum carried by the photon $\gamma$.  The helicity amplitudes
are specified by only one helicity, since parity conservation allows the
independent helicity amplitudes to be defined for $J_\chi \geq \nu \geq 0$ as 
\bea
  B_{\nu'} & \equiv & B_{{\nu'},1} =(-1)^{J_{{\chi}}}B_{-{\nu'},-1}~,
  \nonumber \\
  A_{\nu} & \equiv & A_{{\nu},1} =(-1)^{J_{{\chi}}}A_{-{\nu},-1}~. \nonumber
\eea
Here the second index of the two-index helicities refers to the photon.  To
form the joint angular distribution the $\psipr$ and $\jpsi$ density matrices
must be constructed from the directions of the two electrons forming the
$\psipr$ and the two leptons that decay from the $\jpsi$.\footnote{In $e^+ e^-
\to \gamma^{\ast} \to \psipr$, the polarization of the $\psipr$
along the beam axis is $\pm 1$, so the density matrix giving the polarizations
in the direction of the beam axis (the $z$-axis) is given by $\rho^{(\lambda'
\tilde{\lambda}')} = \epsilon_1^{\ast (\lambda')}
\epsilon_1^{(\tilde{\lambda}')} + \epsilon_2^{\ast (\lambda')}
\epsilon_2^{(\tilde{\lambda}')}$, where $\epsilon^{(\lambda)}$ is the
polarization vector for helicity $\lambda$ defined with components
$\epsilon^{(1)} =  (- 1, - i, 0)/\sqrt{2}$, $\epsilon^{(0)} = (0, 0, 1)$ and
$\epsilon^{(- 1)} = - \epsilon^{(1) \ast} = (1, - i, 0)/\sqrt{2}$.
Generalizing to an arbitrary direction $\hat{n} \equiv (\sin \theta' \cos
\phi', \sin \theta' \sin \phi', \cos \theta')$, we find that the density
matrix $\rho$ for $\psipr$ is
\[
  \rho^{(\lambda', \tilde{\lambda}')} (\theta', \phi') = \sum_{i, j}
  \epsilon_i^{\ast (\lambda')} \epsilon_j^{( \tilde{\lambda}')} 
  \left( \delta^{i j} - n^i n^j \right)~.
\]
Similarly for the $\jpsi$ with $\hat{m}  \equiv (\sin \theta \cos \phi, \sin
\theta \sin \phi, \cos \theta)$, we find the density matrix is
\[
  \rho^{(\lambda, \tilde{\lambda})} (\theta, \phi) = \sum_{i, j}
  \epsilon_i^{\ast (\lambda)} \epsilon_j^{( \tilde{\lambda})} 
  \left( \delta^{i j} - m^i m^j \right)~.
\]}

The angles $\theta'$, $\phi'$ contain information on the polarization of the
$\psi'$, while $\theta$, $\phi$ contain information on the polarization of the
$\jpsi$.  The angle $\theta_{\gamma \gamma'}$, defined by the angle between the
two photons in the $\chi_c$ rest frame, gives information on the necessary
rotation between the two reference frames.  Frames for construction of these
five angles have been shown above in Fig. \ref{fig:angles}.
The joint angular distribution is therefore
\[
  W (\cos \theta', \phi', \cos \theta_{\gamma \gamma'}, \cos \theta, \phi)
  \propto
\]
\beq
  \sum_{\begin{array}{c}
    \nu'  \tilde{\nu}'  ; \mu'  = \pm 1\\[-1.0ex]
    \nu\, \tilde{\nu}\, ; \mu\;   =   \pm 1
  \end{array}} \hspace*{-1em} \rho^{(\mu' - \nu', \mu' -
  \tilde{\nu}')} (\theta', \phi') B_{| \nu' |} B_{| \tilde{\nu}' |} d_{- \nu'
  \nu}^{J_{\chi}} (\theta_{\gamma \gamma'}) d_{- \tilde{\nu}'
  \tilde{\nu}}^{J_{\chi}} (\theta_{\gamma \gamma'}) A_{| \nu |} A_{|
  \tilde{\nu} |} \rho^{\ast (\nu - \mu, \tilde{\nu} - \mu)} (\theta, \phi)~,
  \label{eq:FiveAnglePDF}
\eeq
where $d^{J_{\chi}}_{\nu' \nu}$ are standard Wigner $d$-functions {\cite{PDG}}.

The helicity amplitudes $A_{\nu}$ (with $0 \leq \nu \leq J_\chi$)
are related to the multipole amplitudes $a_{J_\gamma}$ (with 
$1 \leq J_\gamma \leq J_\chi + 1$), using the Clebsch-Gordan coefficients
 $\left\langle j_1, m_1 ; j_2, m_2 |J, M \right\rangle$, by
\beq
  A_{\nu}^{J_{\chi}} = \sum_{J_{\gamma}} 
\sqrt{\frac{2 J_{\gamma} + 1}{2 J_{\chi} + 1}} a_{J_{\gamma}}^{J_{_{\chi}}}  
\left\langle J_{\gamma}, 1 ; 1, \nu - 1 | J_{\chi}, \nu \right\rangle. 
\label{eq:clebsch}
\eeq
This expression leads to the following relationships for the $J_{\chi} = 1$ 
and $J_{\chi} = 2$ cases, respectively:
\bea
  \left(\begin{array}{c}
    A_0^{J = 1}\\
    A_1^{J = 1}
  \end{array}\right) & = & \left(\begin{array}{cc}
    \sqrt{\frac{1}{2}} & \sqrt{\frac{1}{2}}\\
    \sqrt{\frac{1}{2}} & - \sqrt{\frac{1}{2}}
  \end{array}\right) \left(\begin{array}{c}
    a_1^{J = 1}\\
    a_2^{J = 1}
  \end{array}\right)~, \label{eq:clebsch2}\\
   \left(\begin{array}{c}
    A_0^{J = 2}\\
    A_1^{J = 2}\\
    A_2^{J = 2}
  \end{array}\right) & = & \left(\begin{array}{ccc}
    \sqrt{\frac{1}{10}} & \sqrt{\frac{1}{2}} & \sqrt{\frac{2}{5}}\\
    \sqrt{\frac{3}{10}} & \sqrt{\frac{1}{6}} & - \sqrt{\frac{8}{15}}\\
    \sqrt{\frac{3}{5}} & - \sqrt{\frac{1}{3}} & \sqrt{\frac{1}{15}}
  \end{array}\right) \left(\begin{array}{c}
    a_1^{J = 2}\\
    a_2^{J = 2}\\
    a_3^{J = 2}
  \end{array}\right)~.  \label{eq:clebsch3}
\eea
The relationships between $B_{\nu'}$ and $b_{J_{\gamma'}}$ are identical; just
swap all $A_{\nu}$ and $a_{J_{\gamma}}$ with $B_{\nu'}$ and $b_{J_{\gamma'}}$
in Eqs.~(\ref{eq:clebsch})--(\ref{eq:clebsch3}).  These transformation matrices 
are norm-preserving, since the matrices are orthogonal.

\subsection{Quark magnetic moments}

If we define $E1$, $M2$, and $E3$ to be the electric dipole, magnetic quadrupole, and
electric octupole amplitudes, respectively, the magnetic quadrupole amplitudes
are related to the anomalous magnetic moment of the charm quark $\kappa_c$ by
\bea
 a^{J = 1}_2 & \equiv & \frac{M 2}{\sqrt{E 1^2 + M 2^2}} = -
 \frac{E_{\gamma}}{4 m_c} (1 + \kappa_c) \label{eq:a2J1} \\
 a^{J = 2}_2 & \equiv & \frac{M 2}{\sqrt{E 1^2 + M 2^2 + E
 3^2}} = - \frac{3}{\sqrt{5}}  \frac{E_{\gamma}}{4 m_c} (1 + \kappa_c) \\
 b^{J = 1}_2 & \equiv & \frac{M 2}{\sqrt{E 1^2 + M 2^2}} =
 \frac{E_{\gamma'}}{4 m_c} (1 + \kappa_c) \\
 b^{J = 2}_2 & \equiv & \frac{M 2}{\sqrt{E 1^2 + M 2^2 + E 3^2}}
 = \frac{3}{\sqrt{5}}  \frac{E_{\gamma'}}{4 m_c} (1 + \kappa_c) \label{eq:b2J2}
~.
\eea
These expressions are correct
to first order in $E_\gamma/m_c$ or $E_{\gamma'}/m_c$, assuming that the $\psi
(1 S, 2 S)$ are pure $S$ states (no mixing with $D$ states) and that the
$\chi_c$ states are pure $P$ states (no mixing with $F$ states)
\cite{Rosner:2008,SGR:1992}.\footnote{Note the misprint in {\cite{SGR:1992}}
for their equation (41) describing $a_2^{J = 2}$ to first order.  This misprint
was previously noted in footnote 1 of Ref.\ \cite{E835:2002}.}

These first order relationships are derived from the non-relativistic 
interaction Hamiltonian for photon emission from a +2/3 charged quark:
\begin{eqnarray}
  & H_I = & - \frac{e_c}{2 m_c} ( \vec{A}^{\ast} \cdot \vec{p} + \vec{p}
  \cdot \vec{A}^{\ast}) - \mu \vec{\sigma} \cdot \vec{H}^{\ast}
\end{eqnarray}
where $e_c \equiv \frac{2}{3} |e|$, $\mu \equiv (e_c / 2 m_c) (1 + \kappa_c)$,
$\vec{A}^*$ is  the vector potential of the emitted photon, and $\vec{H}^*
\equiv \nabla \times \vec{A}^*$ is the magnetic field of the emitted photon
(both $\vec{A}^*$ and $\vec{H}^*$ are complex conjugated since the photon is
outgoing).  We omit a spin-orbit term of the same order which 
does not affect the $M2$ contribution \cite{KMR:1980,McClary:1983xw}.

The ratios of the predicted multipole amplitudes given by Eqs.\
(\ref{eq:a2J1})--(\ref{eq:b2J2}) are independent of $m_c$ and $\kappa_c$ to 
first order:
\bea
\left( \frac{a^{J = 1}_{2}}{a_2^{J = 2}} \right)_{\rm th} &=&
  \frac{E_{\gamma}^{J = 1}}{E_{\gamma}^{J = 2}}  \frac{\sqrt{5}}{3} = 0.676
  \pm 0.071~,  \label{eq:theory.ratio.1}\\
  \left( \frac{a_2^{J = 1}}{b_2^{J = 1}} \right)_{\rm th} &=& -
  \frac{E_{\gamma}^{J = 1}}{E_{\gamma'}^{J = 1}} = - 2.27 \pm 0.16~,
  \label{eq:theory.ratio.2}\\
  \left( \frac{b_2^{J = 2}}{b_2^{J = 1}} \right)_{\rm th} &=&
  \frac{E_{\gamma'}^{J = 2}}{E_{\gamma'}^{J = 1}} \frac{3}{\sqrt{5}} = 1.000
  \pm 0.015~, \label{eq:theory.ratio.3} \\
  \left( \frac{b_2^{J = 2}}{a_2^{J = 2}} \right)_{\rm th} &=& -
  \frac{E_{\gamma'}^{J = 2}}{E_{\gamma}^{J = 2}} = -0.297 \pm 0.025~.
  \label{eq:theory.ratio.4}
\eea
As the individual amplitudes have corrections of order $(E_\gamma/m_c)^2$, we
conservatively assigned each multipole amplitude a fractional uncertainty equal
to $(E_\gamma/m_c)^2$ (using $m_c = 1.5 \gev$, $\kappa_c = 0$) which was the
dominant source of uncertainty in Eqs.~(\ref{eq:theory.ratio.1})%
--(\ref{eq:theory.ratio.4}).

The $E3$ amplitudes are expected to be small in view of the few-percent admixture 
of the $1^3D_1$ state in the $\psi'$.  Although they are found to be complex 
in Ref.\ \cite{SGR:1992}, we shall include them in fits assuming that they are
real.

\subsection{Lattice QCD predictions}

Dudek {\it et al.} \cite{Dudek:2006,Dudek:2009} performed lattice QCD
calculations for
the charmonium radiative transitions $\chi_{(c1,c2)} \to \gamma \jpsi$.  They
ran lattice simulations at various values of $Q^2$ (the square of the
four-vector of the photon, which is 0 for real photons) and extrapolated to
$Q^2 \to 0$.

For the transition $\chi_{c1} \to \gamma \jpsi$, when extrapolating
the $E1$ and $M2$ amplitudes to $Q^2\to0$ individually, they found that
\beq \label{eqn:chic1lat}
 \frac{M2(Q^2\to0)}{E1(Q^2\to0)} = \frac{-0.020 \pm 0.017}{0.23 \pm 0.03} =
 -0.09 \pm 0.07
\eeq
They concluded that data points at smaller $Q^2$ and improved knowledge of
form factors were needed to make a
meaningful comparison with experimental values \cite{Dudek:2009}.
Similarly for $\chi_{c2} \to \gamma \jpsi$, they found the normalized
multipole amplitudes behaving as $a_2^{J=2} \to -0.39 \pm 0.07$,
$a_3^{J=2} \to 0.010 \pm 0.011$ as $Q^2 \to 0$.

\section{PRIOR EXPERIMENTAL RESULTS \label{sec:prior}}

Tables \ref{pastj1table} and \ref{pastj2table} summarize the results from
previous experiments for $J_{\chi} = 1$ and $J_{\chi} = 2$, respectively.

For transitions involving $\chi_{c1}$, the Crystal Ball experiment at SPEAR
used 921 events of $e^+ e^- \rightarrow \psi' \rightarrow \gamma' \chi_{c1}
\rightarrow \gamma' \gamma J / \psi \rightarrow \gamma' \gamma \ell^+ \ell^-$
to measure both multipole amplitudes. The E-835 experiment used 2090 $p \bar{p}
\to \chi_{c1} \to \gamma J / \psi \to \gamma e^+ e^-$ events to measure the
multipole amplitude $a_2^{J = 1}$. The CLEO-c data sample has $\simeq$40000
$e^+ e^- \rightarrow \psi' \rightarrow \gamma' \chi_{c1} \rightarrow \gamma'
\gamma J / \psi \rightarrow \gamma' \gamma \ell^+ \ell^-$ events after applying
selection criteria.

For transitions involving $\chi_{c2}$, the Crystal Ball experiment used 441
events of $e^+ e^- \rightarrow \psi' \rightarrow \gamma' \chi_{c2} \rightarrow
\gamma' \gamma J / \psi \rightarrow \gamma \gamma' \ell^+ \ell^-$ to measure
both multipole amplitudes. The E-760 and E-835 experiments used 1904 and
5908 $p \bar{p} \to \chi_{c2} \to \gamma J / \psi \to \gamma e^+ e^-$ events,
respectively, to measure the multipole amplitude $a_2^{J = 2}$. The BESII
experiment searched for the multipole amplitudes in a novel method looking at
418 $\psi' \rightarrow \gamma \chi_{c2} \rightarrow \gamma \pi^+ \pi^-$ events
and 303 $\psi' \rightarrow \gamma \chi_{c2} \rightarrow \gamma K^+ K^-$ events,
and the BESII fit also found a value of $b_3^{J = 2} = - 0.027^{+ 0.043}_{-
0.029}$. The CLEO-c data sample has $\simeq$20000 $e^+ e^- \rightarrow \psi'
\rightarrow \gamma' \chi_{c2} \rightarrow \gamma' \gamma J / \psi \rightarrow
\gamma' \gamma \ell^+ \ell^-$ events after applying selection criteria.


\begin{table}[htb]
\caption{Previous experimental values vs.\ theoretical predictions for the
  normalized magnetic quadrupole amplitude for the decays $\chi_{c1}
  \rightarrow \gamma J / \psi$ ($a_2^{J = 1})$ and $\psi' \rightarrow \gamma'
  \chi_{c1}$ ($b_2^{J = 1}$).}
\label{pastj1table}
\begin{center}
  \begin{tabular}{lccc} \hline \hline
  \rule[10pt]{-1mm}{0mm}
    Experiment & $a_2^{J = 1}$ & $b_2^{J = 1}$ & Signal Events \\ \hline
  \rule[12pt]{-1mm}{0mm}
 Crystal Ball {\cite{Oreglia:1980,Oreglia:1982}} & $- 0.002_{- 0.020}^{+ 0.008}$ &
  $0.077_{- 0.045}^{+ 0.050}$ & 921\\
 E-835 {\cite{E835:2002}} & $0.002 \pm 0.032 \pm 0.004$ &  & 2090\\
 Theory ($m_c = 1.5 \gev)$ & $- 0.065 (1 + \kappa_c)$ & $0.029 (1 + \kappa_c)$ 
  & \\ \hline \hline
  \end{tabular}
\end{center}
\end{table}


\begin{table}[htb]
  \caption{Previous experimental values vs.\ theoretical predictions for the
normalized magnetic quadrupole amplitudes for the decays $\chi_{c2} \rightarrow
\gamma J / \psi$ ($a_2^{J = 2})$ and $\psi' \rightarrow \gamma' \chi_{c2}$
($b_2^{J = 2}$).
  \label{pastj2table}}
\begin{center}
  \begin{tabular}{lccc} \hline \hline
  \rule[11pt]{-1mm}{0mm}
    Experiment & $a_2^{J = 2}$ & $b_2^{J = 2}$ & Signal Events\\ \hline
  \rule[12pt]{-1mm}{0mm}
  Crystal Ball {\cite{Oreglia:1980,Oreglia:1982}} & $- 0.333_{- 0.292}^{+ 0.116}$ &
  $0.132_{- 0.075}^{+ 0.098}$ & 441\\
  E-760 {\cite{E760:1993}} & $- 0.14 \pm 0.06$ &  & 1904\\
  E-835 {\cite{E835:2002}} & $- 0.093_{- 0.041}^{+ 0.039} \pm 0.006$ &  &  5908\\
   BESII {\cite{BES:2004}} &  & $- 0.051_{- 0.036}^{+ 0.054}$ & 731\\
   Theory ($m_c = 1.5 \gev)$ & $- 0.096 (1 + \kappa_c)$ & $0.029 (1 +
   \kappa_c)$ & \\ \hline \hline
  \end{tabular}
\end{center}
\end{table}

Many of these experimental results disagreed with theory which predicted ratios
given in Eqs.~(\ref{eq:theory.ratio.1})--(\ref{eq:theory.ratio.4}). The ratios
of the averages of previous experimental values compared with theory values
are
\begin{eqnarray}
  \left( \frac{a_2^{J = 1}}{a_2^{J = 2}} \right)\tsb{exp} = \frac{-
  0.002 \pm 0.020}{- 0.13 \pm 0.05} = ~~ 0.02^{+0.17}_{-0.16}
& \stackrel{?}{=} & \left( \frac{a_2^{J = 1}}{a_2^{J = 2}} \right)\tsb{th}
 = 0.676 \pm 0.071 \\
  \left( \frac{a_2^{J = 1}}{b_2^{J = 1}} \right)\tsb{\exp} =
  \frac{- 0.002 \pm 0.020}{0.077 \pm 0.050} = -0.02^{+0.30}_{-0.32}
& \stackrel{?}{=} &\left( \frac{a_2^{J = 1}}{b_2^{J = 1}} \right)\tsb{th}
 = - 2.27 \pm 0.16 \\
  \left( \frac{b_2^{J = 2}}{b_2^{J = 1}} \right)\tsb{\exp}\, =~~
  \frac{0.132 \pm 0.075}{0.077 \pm 0.050} =~~~ 1.5^{+2.2}_{-1.1}~~
& \stackrel{?}{=} & \left( \frac{b_2^{J = 2}}{b_2^{J = 1}} \right)\tsb{th}
 = 1.000 \pm 0.015 \\
  \left( \frac{b_2^{J = 2}}{a_2^{J = 2}} \right)\tsb{\exp}\,  =~~
  \frac{0.132 \pm 0.075}{-0.13 \pm 0.05} = -1.01^{+0.60}_{-0.93}
& \stackrel{?}{=} &  \left( \frac{b_2^{J = 2}}{a_2^{J = 2}} \right)\tsb{th}
 = -0.297 \pm 0.025~.
\end{eqnarray}
The first two ratios which involve the multipole amplitudes that have the most 
statistical significance strongly disagree with their theoretical predictions.
As the ratios are independent of $m_c$, $\kappa_c$ and any specific quarkonium
potential model to first order in $E_\gamma/(4 m_c)$, we expect good agreement
between theory and experiment.

\section{DETECTOR, DATA, AND MONTE CARLO \label{sec:detdtmc}}

\subsection{The CLEO detector}

Data were acquired at the $\psip$ resonance at $\sqrt{s} = 3.686 \gev$
using the CLEO-c detector located at the Cornell Electron Storage Ring (CESR),
a symmetrical $e^+ e^-$ collider \cite{Kubota:1991ww,Peterson:2002sk}.  The
solid angle for detecting both charged and neutral particles is 93\% of $4\pi$.
The photons were detected as showers in a CsI (Tl) calorimeter consisting
of 7784 crystals, which achieved a photon energy resolution of 2.2\% at
$1 \gev$ and 5\% at $100 \mev$. The azimuthal and polar angular
resolution for $100 \mev$ photons is $\sigma_{\phi_{\rm azim}} \approx 11\;
\text{mrad}$ (19 mrad) and $\sigma_{\theta_{\rm polar}} \approx 0.8
\sigma_{\phi_{\rm azim}} \sin\theta_{\rm polar}$ (10 mrad) in the barrel
(endcap) region of the crystal calorimeter. Charged particles were detected
using a set of two cylindrical drift
chambers enclosed within a superconducting solenoid with a 1.0 T magnetic
field directed along the beam axis. The outer drift chamber achieved a
momentum resolution of $\approx$$0.6\%$ at $p = 1 \gev$ and an azimuthal
and polar angular resolution of $\sigma_{\phi_{\rm azim}} \approx 1\;
\text{mrad}$ and $\sigma_{\theta_{\mathrm{polar}}} \approx 4\; \text{mrad}$
\cite{Kubota:1991ww}. (In this article, 
$c=1$ in mass and momentum units.) The inner
six-layer stereo drift chamber is used to accurately measure the location of
charged particles along the beam axis.

\subsection{Data sets and expected number of events}
For our analysis, we used the recent CLEO-c data set taken at the $\psip$
events consisting of a sample of $(24.45 \pm 0.49) \times 10^6$ $\psip$
events with a total luminosity of 48.07/pb \cite{heltsley:2008kb}.
Using known branching fractions \cite{PDG} and the known sizes of the CLEO data
sample, we can expect that $91900 \pm 6600$ $J_\chi = 1$ signal events and
$48200 \pm 3600$ $J_\chi =2$ signal events are originally present in the data
sample.

\subsection{Phase space Monte Carlo}\label{phsp}

For each of the decay sequences ($J_\chi = 1$ and $J_\chi=2$), a 4.5 million
event phase space Monte Carlo (MC) data sample was generated. The phase space
MC was generated with EvtGen \cite{Lange:2001uf} with final state radiation 
simulated with PHOTOS \cite{Barberio:1993qi}.

The purposes of the phase space Monte Carlo are threefold. First, it is used to
account for the variable angular efficiency of the detector after selection
criteria have been applied, when performing the maximum likelihood fit (see
Sec.~\ref{fitting_procedure}). Second, the phase space MC events are used to
simulate signal MC with non-zero multipole amplitudes, $a_2, b_2$ (and $a_3,
b_3$ for $J_{\chi} = 2$) via the rejection method. This is achieved by taking
the five angles $\theta', \phi', \theta_{\gamma \gamma'}, \theta, \phi$ for
each phase space event and calculating the probability of that event occurring
at those angles for the PDF $\PDFz$ with the input values of the multipole
amplitudes $\pmbA_0$. The probability for the event occurring at that angle
is then compared to a random number uniformly distributed between 0 and 1.
Then, our simulated signal MC obeying the PDF $\PDFz$ consists of the events
that are more probable than the corresponding random numbers.  The third
purpose of the phase space Monte Carlo is to generate projections to overlay
upon histograms of data values. For example, after a fit to data extracts
values of $a_2, b_2$ for a $J_{\chi} = 1$ fit, the phase space MC can be used
to generate projections in the five angles with the fitted values of $a_2, b_2$
to be compared with the data.

\subsection{Generic Monte Carlo}

In order to properly simulate feed-across into the selected data sample from
non-signal $\psi^\prime$ decays, a ``generic'' MC sample was prepared.  This
sample consists of approximately 120 million $\psi^\prime$ decays, using our
best estimate for all measured branching fractions 
\cite{heltsley:2008kb,PDG,Bai:2001ct,Athar:2004dn,Pedlar:2005px,adams:2008ab,
Naik:2008dk,Ecklund:2008hg,He:2008vna,Mitchell:2008fb,Asner:2008nw}
and matrix elements for the decays of $\psi^\prime$ and its decay
products; unmeasured hadronic decays are simulated using JETSET 
\cite{Sjostrand:2000wi}.
The signal events ($\psi' \rightarrow \gamma' \chi_{(c 1, c 2)} \rightarrow
\gamma' \gamma J / \psi$) were replaced with phase space MC events selected to
have the desired $a_2$ and $b_2$ admixture (via the rejection method as
described in Sec.\ \ref{phsp}).

\section{SELECTION CRITERIA \label{selection_criteria}}

Tuning of the selection criteria was designed to eliminate non-signal
``impure'' background events, while
selecting the largest number of signal events.  For kinematic regions in which
it was uncertain how to apply selection criteria, we attempted to minimize the
quadrature sum of the statistical uncertainty from signal events and the
systematic uncertainty from impure events.  Many of the starting points for our
selection criteria are taken from a CLEO-c study
\cite{heltsley:2008kb} of $\psip \to h + \jpsi$
branching fractions that included our signal decays.

All tracks and showers investigated are required to pass standard CLEO-c 
criteria prior to any attempts at kinematic fitting. For tracks, we ensure 
that the track originated from near the interaction point ($r_0 < 2$~cm 
and $|z_0 - z_{\mathrm{i.p.}}| < 10$~cm), is from the well-modeled region 
of the barrel ($|\cos\theta_{\rm polar}| < 0.83$) or the endcap 
($0.85 < |\cos\theta_{\rm polar}| < 0.93$), 
and has a momentum between 1\% ($18.4 \mev/c$) and 120\% ($2.21 \gev/c$) of 
the beam momentum. The requirement for a shower is that it is not matched 
to a track, has $|\cos\theta_{\rm polar}| < 0.79$ or 
$0.85 < |\cos\theta_{\rm polar}| < 0.93$, and has an energy between 1\% 
and 120\% of the beam energy.

\begin{figure}[htb]
\begin{center}
\includegraphics[width=0.91\textwidth]{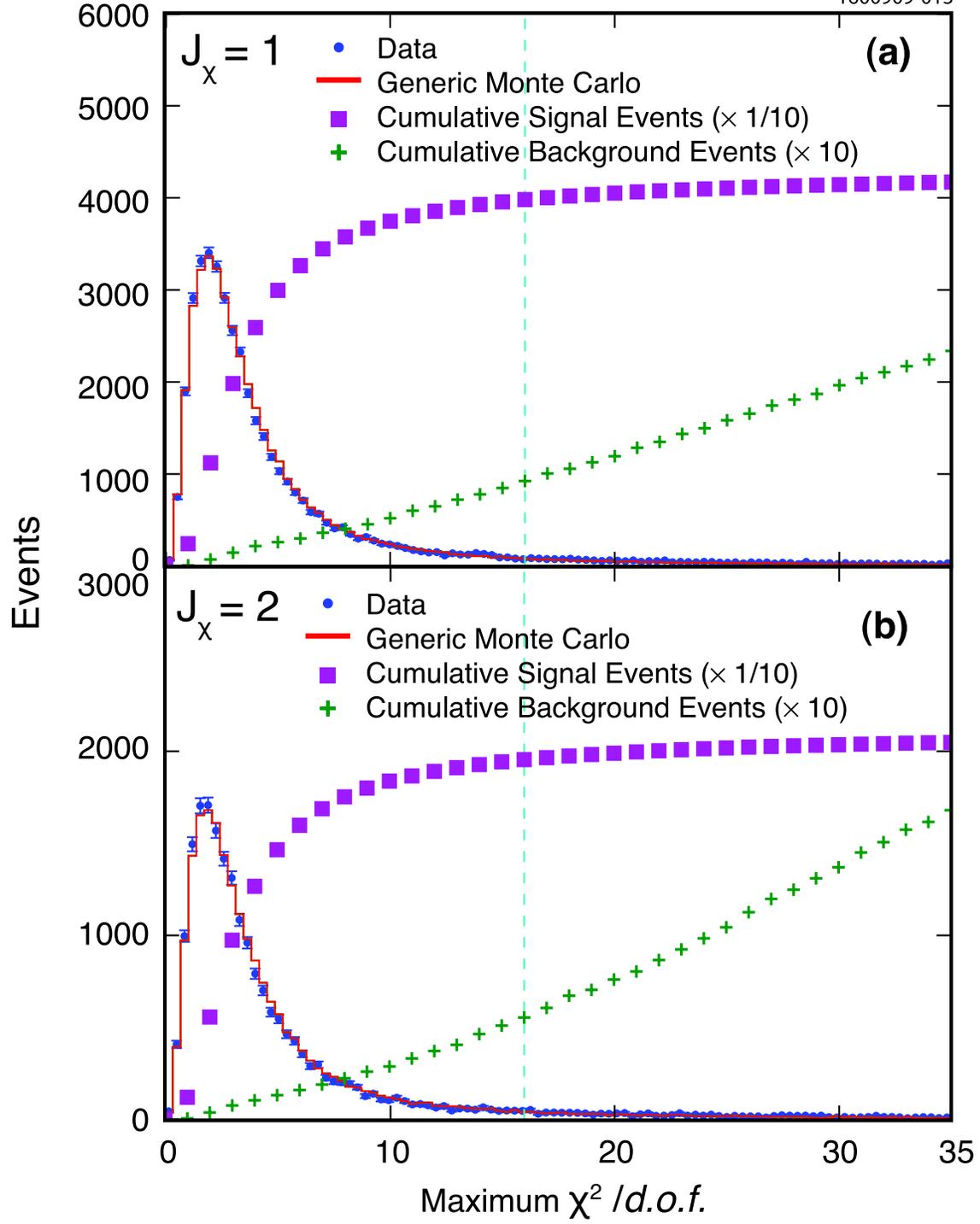}
\end{center}
  \caption{Maximum reduced $\chi^2$ in all kinematic fits (including
  vertex fits) in generic Monte Carlo and data. Events with a maximum
  reduced $\chi^2$ below 16 (the dashed vertical line) are kept.  Cumulative
  totals for the number of signal and impurity background events are also
  plotted for each potential value of a maximum reduced $\chi^2$.  (a)
  $J_\chi = 1$ and (b) $J_\chi = 2$.
\label{fig:maxredchisq}}
\end{figure}

All candidate events require at least two tracks and two showers to be
identified. The two tracks and two showers used (if more are present) will be
those with the greatest energies. Two kinematic fits are then performed to
generate the four 4-vectors used in the analysis. First, a 1C kinematic fit to
the $J / \psi$ mass is performed starting with the two tracks, allowing
shower(s) identified as bremsstrahlung photons to be associated with a track.
Bremsstrahlung photons are identified if a shower that is not matched to a track
is located within 100 mrad of the initial momentum vector of a track.
If bremsstrahlung photons are
identified, the lepton four-vector used is the sum of the kinematically fit
4-vectors of the lepton plus all associated bremsstrahlung photons. Second, a
4C kinematic fit to the $\psip$ 4-vector is performed and the result of this fit
is then subjected once more to the original 1C fit. The $\psip$ 4-vector
is calculated from the angle at which the electron and positron beams
intersect (4 mrad) and the beam energy of the given
run.  For both the 1C and 4C kinematic fits, we require the reduced $\chi^2$
for both the vertex and kinematic fit to be less than 16 as shown in Fig.\
\ref{fig:maxredchisq}.  This value was found by minimizing the quadrature sum
of the impurity systematic uncertainty and statistical uncertainty.

\begin{figure}[htb]
\begin{center}
\includegraphics[width=0.80\textwidth]{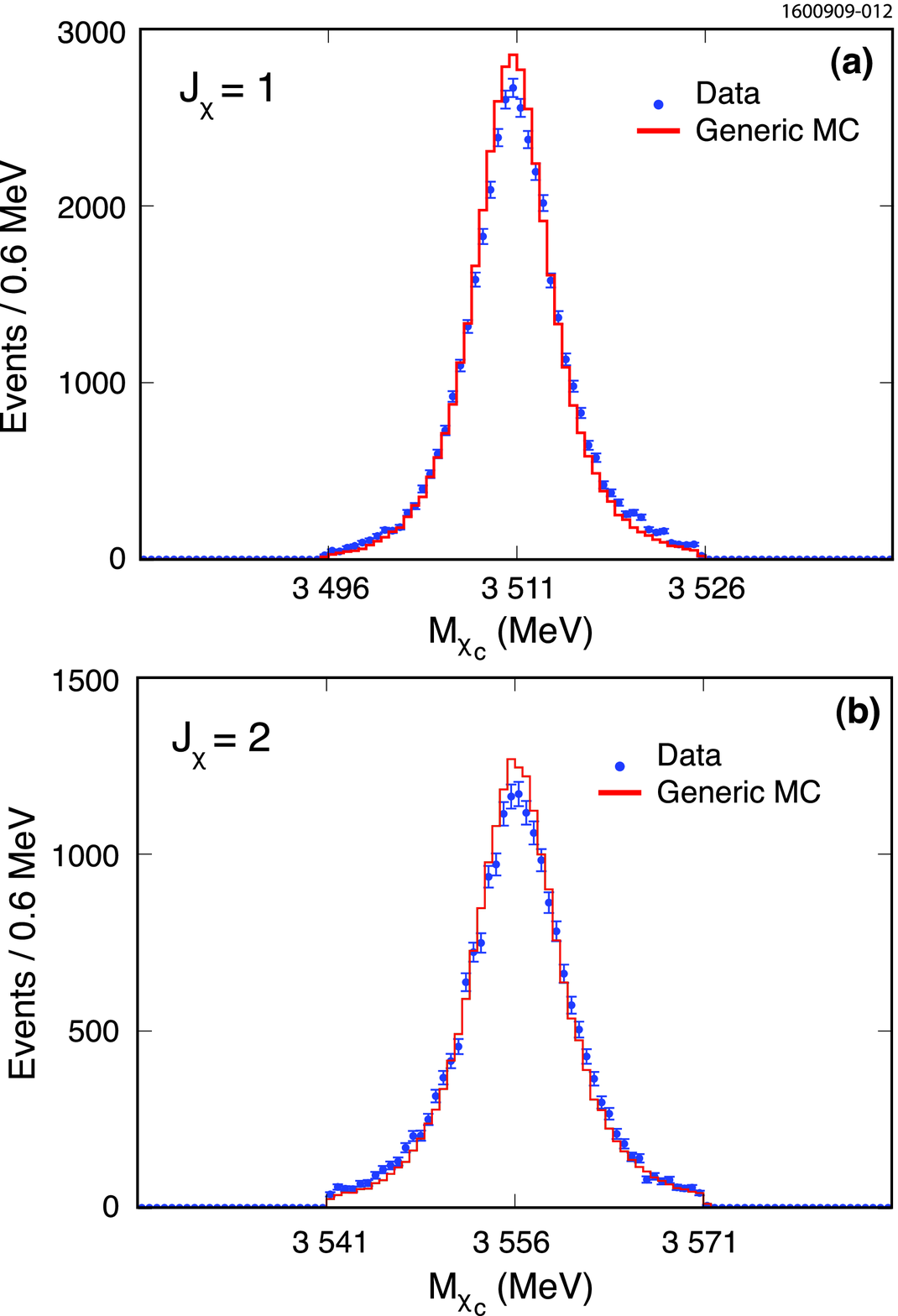}
\end{center}
\caption{Plot of the $\chi_{cJ}$ mass as calculated from subtracting the
four-vector
of the $\gamp$ from the $\psip$ four-vector.  This variable is not used
as a selection criterion because the 1C and 4C kinematic fits ensure that this
criterion is redundant with the $\chi_{c1}$ mass selection criterion generated 
by adding the $\jpsi$ and $\gamu$ four-vectors.  (a) $J_\chi = 1$ and (b) 
$J_\chi = 2$.
\label{fig:chic_mass_alt}}
\end{figure}

To identify signal events through the $J_{\chi}$ radiative cascade,
we require the reconstructed $\chi_{cJ}$ mass to be within 15 MeV of
the true $\chi_{cJ}$ mass as constructed by adding the $\jpsi$ and
$\gamma$ four-vectors together:
\beq \nonumber
  m_{\chi_{cJ}} = \sqrt{|p_{\jpsi} + p_{\gamma} |^2}\\
=\sqrt{|p_{\ell^+} + p_{\ell^-} + p_{\gamma}  |^2}
\eeq
We do not apply a selection criterion based on the other $\chi_{cJ}$ mass 
reconstructed from the $\psip$ and $\gamma'$, as the 4C kinematic fit ensures
that this criterion is redundant (see Fig.~\ref{fig:chic_mass_alt}).

Signal events must also have the $J / \psi$ decay to $e^+ e^-$ or $\mu^+
\mu^-$, so we require the two tracks to be well-identified as both being
electrons or muons. We achieve this by looking at the ratio of the
energy deposited in the calorimeter to the momentum of the track ($E / p$).
We identify both tracks as electrons if the smaller $E / p$ ratio is greater
than 0.5 and the larger $E / p$ ratio is greater than 0.85. Similarly, we
identify both tracks as muons if $(E / p)_{\rm smaller} < 0.25$ and
$(E / p)_{\rm larger} < 0.5$.  This results in a clean $e$--$\mu$ separation.

To restrict major sources of background, we apply additional criteria to 
address the modes with large branching fractions:
\bea
\mathcal{B} (\psi' \to \pi^0 \pi^0 \jpsi) &=& (16.84 \pm 0.33)\% \nonumber \\
\mathcal{B} (\psi' \to \eta \jpsi) &=& (3.16 \pm 0.07)\% \nonumber \\
\mathcal{B} (\psi' \to \pi^0 \jpsi) &=& (1.26 \pm 0.13) \times 10^{- 3}
\eea
The dominant background mode $\psi' \to \pi^0 \pi^0 \jpsi \to \gamma \gamma
\gamma \gamma \ell^+ \ell^-$ is reduced by requiring the third most energetic
shower in the event (excluding those photons identified as bremsstrahlung
photons) to have an energy of less than 30 MeV. To reduce the contributions of
the background modes with monochromatic $\jpsi$ momentum,
 $\psi' \to \eta \jpsi \to \gamma \gamma \jpsi$ and $\psi' \to \pi^0
\jpsi \to \gamma \gamma \jpsi$ where $p (\jpsi) |_{\psi' \to \eta \jpsi} =
199 \mev$ and $p (\jpsi) |_{\psi' \to \pi^0 \jpsi} = 528 \mev$,
we require the $\jpsi$ momentum to lie between $240 \mev$ and $510 \mev.$
Note that the signal transition generates no events with a $\jpsi$ momentum
below 238 MeV (318 MeV for $J_{\chi} = 2$) or above 542 MeV.
\bigskip


\section{FITTING THE DATA \label{sec:fits}}

\subsection{Basic approach and procedure} \label{fitting_procedure}

We find the multipole amplitudes by performing a maximum likelihood fit of the
selected data events to the probability distribution function (PDF) $\PDF$
given by Eq.~(\ref{eq:FiveAnglePDF}). Events are selected according to the
criteria described in Sec.~\ref{selection_criteria} and each event is
described by a set of five angles $\Omega \equiv (\theta', \phi',
\theta_{\gamma \gamma}, \theta, \phi)$ defined in Fig.~\ref{fig:angles}. The
PDF $W (\Omega ; \pmbA)$ gives the probability for an event with angles
$\Omega$ to occur given a set of multipole amplitudes $\pmbA \equiv
(a_i, b_j)$. The PDF in Eq.~(\ref{eq:FiveAnglePDF}) is written in terms
of helicity amplitudes, but can be written in terms of multipole amplitudes as
$W (\Omega ; \pmbA)$ using Eq.~(\ref{eq:clebsch}).
The total likelihood for $N_d$ data events to be described by $\PDF$ is
\beq
 \mathcal{L}_W(\pmbA) \equiv \prod^{N_d}_{d = 1} W (\Omega_d ; \pmbA).
\label{eq:likelihood_W}
\eeq

The initially unknown angular detector efficiency $\epsilon (\Omega)$ describes
the probability that an event occurring at the angles $\Omega$ will be
registered by the detector and pass the selection criteria.  We define a new
normalized PDF to account for this detector efficiency $\epsilon (\Omega)$:
\begin{eqnarray}
  F (\Omega ; \pmbA) \equiv \frac{\epsilon (\Omega) W (\Omega ;
  \pmbA)}{\int \epsilon (\Omega') W (\Omega' ; \pmbA) d \Omega'} &
  &  \label{eq:detectorPDF}
\end{eqnarray}
and note that the original PDF $W (\Omega ; \pmbA)$ is of the form
\begin{eqnarray}
  W (\Omega ; \pmbA) = \sum_{i j k l} a_i a_j b_k b_l G_{i j k l}
  (\Omega).
\end{eqnarray}
The functions $G_{i j k l} (\Omega)$ are obtained from the expression for
$\PDF$, so this form allows separation of the parameters being determined
(the
multipole amplitudes $\pmbA$) and the data points (the angles $\Omega$). This
allows us to write the denominator of the PDF in Eq.~(\ref{eq:detectorPDF}) as
\begin{eqnarray*}
  \int \epsilon (\Omega') W (\Omega' ; \pmbA) d \Omega' & = & \int
  \epsilon (\Omega') \sum_{i j k l} a_i a_j b_k b_l G_{i j k l} (\Omega') d
  \Omega'\\
  & = & \sum_{i j k l} a_i a_j b_k b_l  \int \epsilon (\Omega') G_{i j k l}
  (\Omega') d \Omega'\\
  & = & \sum_{i j k l} a_i a_j b_k b_l I_{i j k l}
\end{eqnarray*}
where the detector-efficiency-dependent integrals $I_{i j k l} \equiv \int
\epsilon (\Omega') G_{i j k l} (\Omega') d \Omega'$ are independent of the
fitting parameters $\pmbA$. The integrals $I_{i j k l}$ can be
approximated by a Monte Carlo numerical integration technique. Using a large
sample of phase space Monte Carlo events (Sec.~\ref{phsp}) generated
uniformly in the five angles $(\cos \theta', \phi', \cos \theta_{\gamma
\gamma'}, \cos \theta, \phi)$, we record whether each phase space MC event is
reconstructed and passes the selection criteria. Using the known angular
functions $G_{i j k l} (\Omega)$, we approximate the integral $I_{i j k l}$ as
\begin{eqnarray}
  I_{i j k l} & \equiv & \int \epsilon (\Omega') G_{i j k l} (\Omega') d \Omega' 
\nonumber \\
  & \cong & \frac{1}{N_{\rm phsp}} \sum^{N_{\rm phsp}}_{p = 1} \Theta
 (p) G_{i j k l} (\Omega_p) \label{eq:integral_approx} \label{eq:phsp_integrals}
\end{eqnarray}
where $\Theta (p)$ is 1 (0) if the $p$th phase space event is (not)
reconstructed and $N_{\rm phsp}$ is the total number of phase space events.

The most likely form of the parameters $\pmbA$ given the PDF $F(\Omega; \pmbA)$
is found by maximizing the logarithm of the likelihood, which is given by 
Eq.(\ref{eq:likelihood_W})
with the PDF $F$ instead of $W$.  The logarithm of the likelihood that the
parameters $\pmbA$ in the PDF $F (\Omega ; \pmbA)$ describe the $N_d$ data
events occurring at angles $\Omega_d$ is
\begin{eqnarray}
  \log \mathcal{L}(\pmbA) &\equiv& \log \prod_{d=1}^{N_d} F(\Omega_d; \pmbA) =
  \sum^{N_d}_{d=1} \log F (\Omega_d ; \pmbA)
  \nonumber\\
  & = & \sum^{N_d}_{d=1} \left[ \log \epsilon (\Omega_d) + \log W (\Omega_d ;
  \pmbA) - \log \sum_{ijkl} a_i a_j b_k b_l I_{i j k l} \right] .
\end{eqnarray}
The first term in $\log \mathcal{L}$ is independent of the $\pmbA$, so the
log likelihood only depends on the detector efficiency through the phase space
integrals.  We reduce the number of parameters in the fit by recognizing that
the multipole amplitudes are normalized (\eg, $a_1^2 + a_2^2 + a_3^2 = 1$).
This method of performing an unbinned maximum likelihood over an angularly
varying detector efficiency was first developed in Ref.\ \cite{Cassel:1981}.
The multi-dimensional optimization of $\log \mathcal{L}'(\pmbA)$ was achieved
using the {\sc Minuit Migrad} variable-metric fitting routine \cite{minuit}.

\subsection{Statistical results of five-angle fits} \label{five_angle_fits}

\subsubsection{$J_{\chi} = 1$ fits}
\begin{figure}[htb]
\begin{center}
\includegraphics[width=0.77\textwidth]{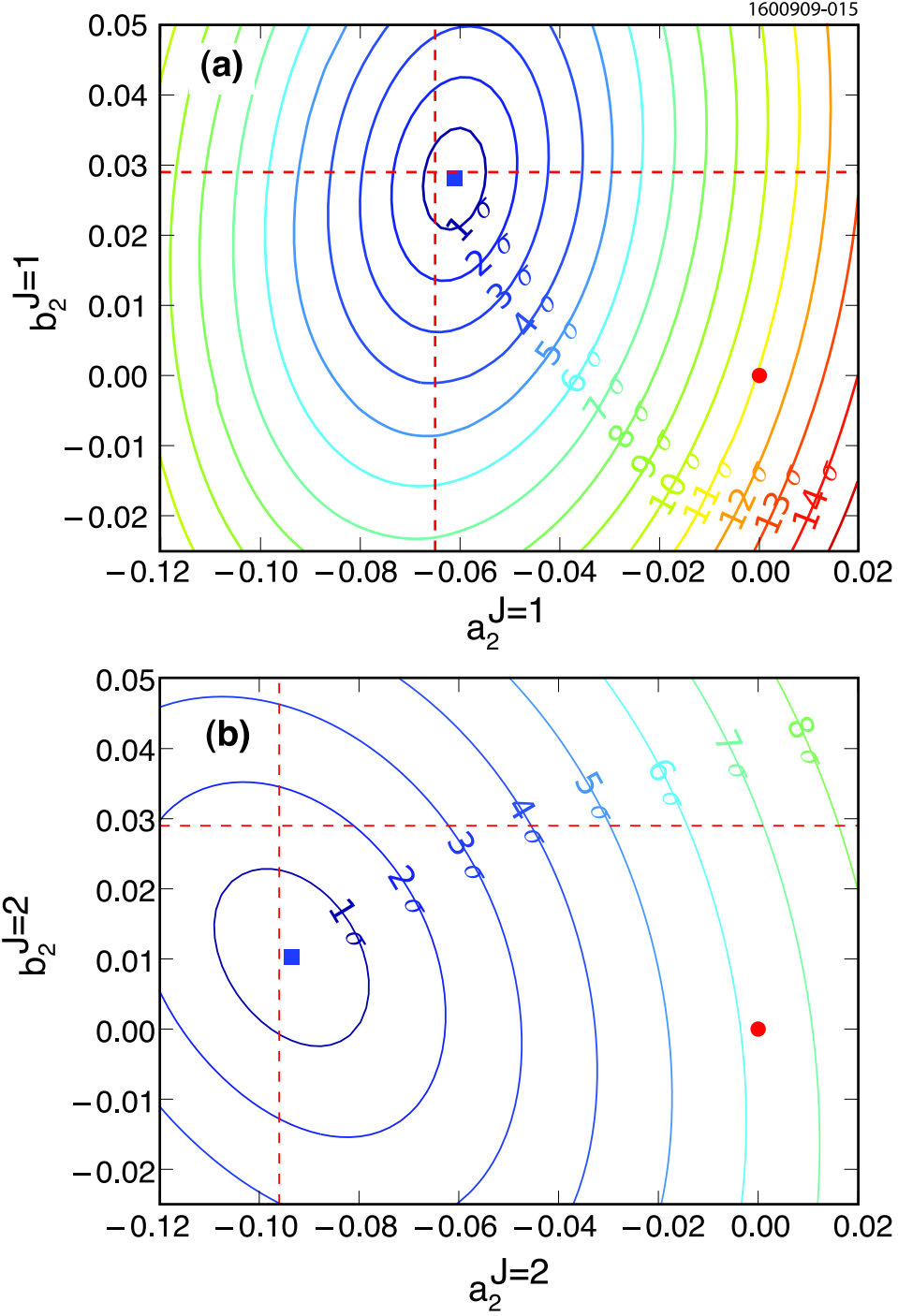}
\end{center}
\caption{(a) $J_\chi = 1$ and (b) $J_\chi = 2$ log likelihood contours
as functions of $(a_2,b_2)$ for two-parameter fits.  The fitted values (the
solid squares) are $(a_2, b_2) = (-0.0611, 0.0281)$ for $J_\chi = 1$ and 
$(a_2, b_2, a_3, b_3) = (-0.093, 0.010, 0,0)$ for $J_\chi = 2$.  These are,
respectively, $11.1 \sigma$ and $6.2 \sigma$ from pure $E1$ (the solid circles).
The theoretical values to first order in $E_\gamma/m_c$ with $\kappa_c = 0$ 
are given by the dashed lines.}
\label{fig:contour}
\end{figure}

The result of the two-parameter fit to the $J_{\chi}=1$ data is $a_2^{J=1} =
-0.0611 \pm 0.0063$, $b_2^{J=1} = 0.0281 \pm 0.0073$, based on 39363 events.
The efficiency integrals in the denominator were calculated by simulating 4.5
million phase space MC events taking account of the detector geometry and
selection criteria; 39.6\% of events were reconstructed. Contours are shown in
Fig.~\ref{fig:contour}(a) of $\sqrt{2 \Delta \log {\cal L}}$, where
$\Delta \log {\cal L}$ is the difference in log likelihood between the
fitted values of $a_2,b_2$ and any other values.  For a pure $E1$ transition
($a_2=b_2=0$) the value of $\chi_{E1} \equiv \sqrt{2 \Delta \log {\cal L}}$
is $11.1$.


The projections of the data in each of the five angles may be compared with
curves based on a pure $E1$ distribution and on the fitted $M2/E1$ admixture.  The
angle $\theta$ is of particular importance as it is the angle that most clearly
shows the preference of the data for an $M2/E1$ admixture over a pure $E1$
transition. The projection for $\costhp$ also shows slightly better agreement
with data with the fitted $M2/E1$ admixture than with a pure $E1$ transition.
For the 50-bin histograms in $\costh$, the reduced $\chi^2$
($\chi^2/N\tsb{d.o.f.}$)~\footnote{The number of degrees of
freedom $N\tsb{d.o.f.} = N\tsb{bins} - N\tsb{params} - 1$ where $N\tsb{bins}$
is the number of bins in the histogram, and $N\tsb{params}$ is the number
of free parameters in the fit.  The minus one accounts for the fact that the
projections are normalized to contain the same number of events as the original
data set.} comparing the data with the projection at the fitted
values is $42.7/47=0.91$, while data and the pure $E1$ projection have a
$\chi^2/N\tsb{d.o.f.}$ of $108.5/49=2.21$.
 
\begin{figure}[htb]
\begin{center}
\includegraphics[width=0.75\textwidth]{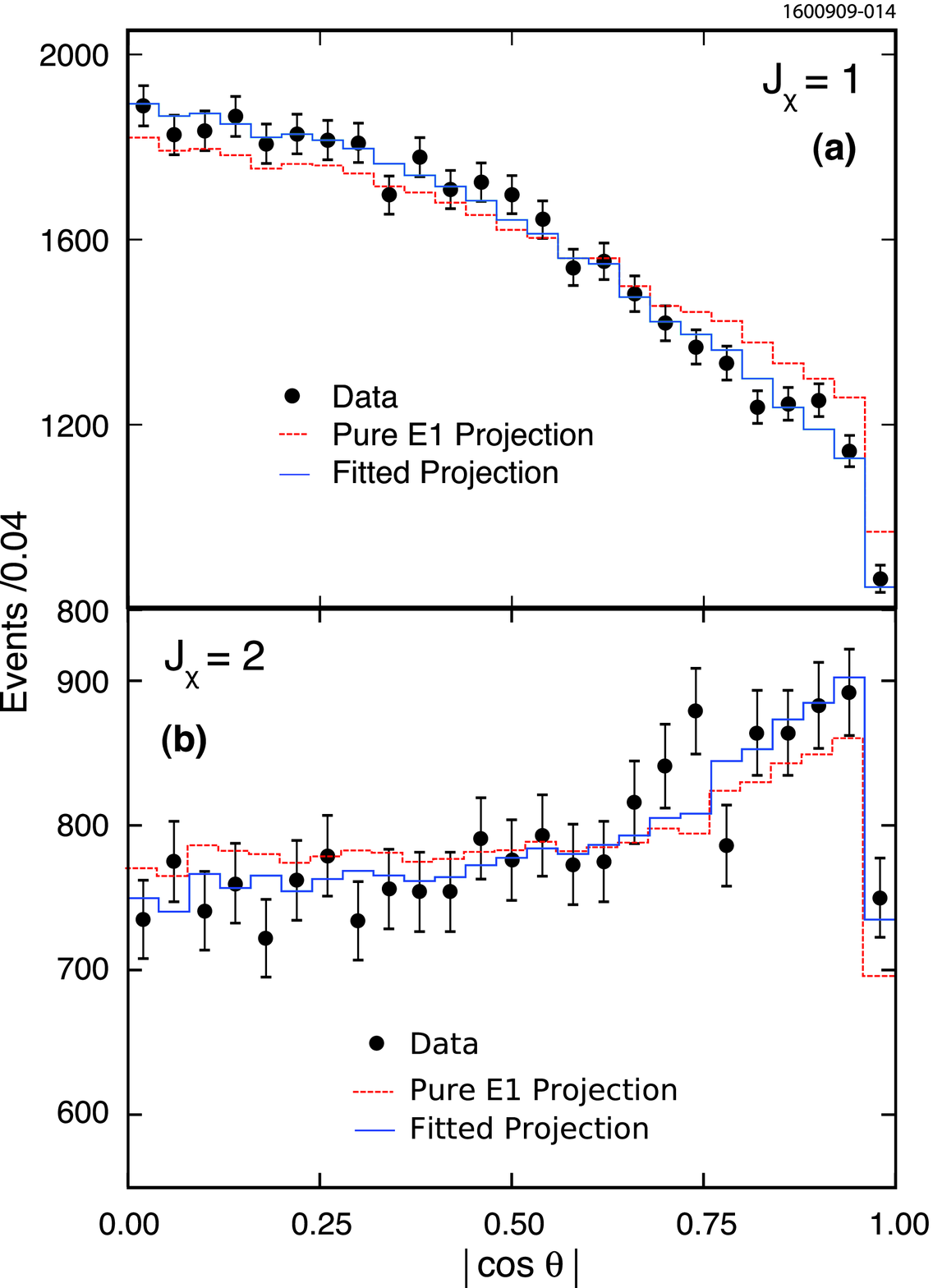}
\end{center}
\caption{(a) $J_\chi = 1$ and (b) $J_\chi = 2$ projections of $\cos
\theta$ after using parity transformations to fold the data set into positive
$\costhp, \phip, \costhgg, \costh$.  For $J_\chi = 1$ ($J_\chi = 2$) the values
of $\chi^2/N\tsb{d.o.f.}$ for the 25 bin histogram describing the data to
correspond with the two-parameter $(a_2,b_2)$ fitted projection are $16.2/22 =
0.74$ ($20.3/22 = 0.92$) and to correspond with the pure $E1$ projection are
$80.29/24=3.35$ ($35.5/24=1.48$).  The fitted and pure $E1$ projections are
selected from the same phase space MC sample (via the rejection method),
resulting in the correlation of statistical fluctuations in the two
projections.}
\label{fig:half_scale_costheta_proj}
\end{figure}

Using the parity transformations described in Ref.\ \cite{Oreglia:1982}, we can 
fold four of the five angles into the positive domain without modifying the
value of the likelihood calculated through $\PDF$.  In Fig.\
\ref{fig:half_scale_costheta_proj}(a) we show that the data are
well-matched with the projection in $|\costh|$ with the fitted values of
$\pmbA$, but poorly matched with the pure-$E1$ $|\costh|$ projection.

When we fix the ratio of the parameters to the theoretical ratio, given by Eq.\
(\ref{eq:theory.ratio.2}), $a_2^{J=1}/b_2^{J=1} = -2.274$, we can perform a
one-parameter fit to the five-angle $J_{\chi}=1$ data set.  The result of this
one-parameter fit is $a_2^{J=1} = -0.0615 \pm 0.0055$, $b_2^{J=1} =
-a_2^{J=1}/2.274 = 0.0271 \pm 0.0024$, with a value of $\chi_{E1} = \sqrt{2
\Delta \log {\cal L}} = 11.1$) nearly identical to the results of the
two-parameter fit.  The results of these two fits are compared in Table
\ref{tab:fit1}.


\begin{table}[htb]
\caption{$J_{\chi}=1$ five-angle fit results.  The fits were performed on 39363
events satisfying the selection criteria described in Sec.\
\ref{selection_criteria}.  $\chi_{E1} \equiv \sqrt{2\Delta \log
\mathcal{L}}$ is the number of standard deviations by which the fitted value
differs from the pure $E1$ value.
\label{tab:fit1}}
\begin{center}
\begin{tabular}{lcccccc} \hline \hline
  \rule[12pt]{-1mm}{0mm}
Fit & $a_2^{J=1}$ & $b_2^{J=1}$ & $\chi_{E1}$ \\
 & $(10^{-2})$ & $(10^{-2})$ & \\ \hline
   Two-parameter & $-6.11 \pm 0.63$ &  $2.81 \pm 0.73$ & 11.1 \\
   One-param.\ ($a_2/b_2 = -2.274$) & $-6.15 \pm 0.55$ & $2.71 \pm 0.24$ & 11.1 \\
   Theory ($m_c = 1.5 \gev$) & {~~$-6.5(1 + \kappa_c)$~~}
 & {~~$2.9(1+\kappa_c)$~~} & \\ \hline \hline
\end{tabular}
\end{center}
\end{table}

\subsubsection{$J_{\chi} = 2$ fits}
As the $J_{\chi}=2$ PDF is parameterized by four multipole amplitudes
($a_2,b_2,a_3,b_3$), there are several choices of fits to be performed.
The simplest would be a two-parameter fit with $a_3=b_3=0$, as the $E3$
amplitudes should be zero in the absence of significant S--D state mixing.
For this type of fit to the
19755 signal events, we find $a_2^{J=2} = -0.093
\pm 0.016$, $b_2^{J=2} = 0.010 \pm 0.013$ with these fit values favored
by $6.2 \sigma$ over a fit with pure $E1$.

Allowing for S--D mixing in the $\psip$ state, the $b_3^{J=2}$ amplitude may
be non-zero.  When we perform a three-parameter fit (setting $a_3^{J=2}$ = 0),
we find $a_2^{J=2}=-0.093 \pm 0.016$, $b_2^{J=2} = 0.007 \pm 0.013$, $b_3^{J=2}
= -0.008 \pm 0.011$, favored by $6.3 \sigma$ over pure $E1$.

When we allow a non-zero $b_3^{J=2}$ amplitude, but fix the ratio of
$a_2^{J=2}/b_2^{J=2}=-3.367$ by Eq.~(\ref{eq:theory.ratio.4}), we can
perform a two-parameter fit that allows for $S-D$ mixing in the $\psip$ state.
The results of this two-parameter fit are $a_2^{J=2} = -0.092 \pm 0.016$,
$b_2^{J=2} = -a_2^{J=2}/3.367 = 0.027 \pm 0.005$, $b_3^{J=2} = -0.001 \pm
0.011$, favored by $6.1 \sigma$ over pure $E1$.

When we perform the fit for the full four parameters ($a_2, b_2, a_3, b_3$), we
find $a_2^{J=2} = -0.079 \pm 0.019$, $a_3^{J=2} = 0.002 \pm 0.014$,
$b_2^{J=2} = 0.017 \pm 0.014$, $b_3^{J=2} = -0.008 \pm 0.012$, favored by
$6.4 \sigma$ over pure $E1$.

For the five-angle fit with two parameters, we plot the data with the pure $E1$
projection and the fitted value projection of $|\costh|$ in Fig.\
\ref{fig:half_scale_costheta_proj}(b).  As for $J_\chi = 1$, the
fitted values match the data better than the pure $E1$ projection.

The results of the above fits are summarized in Table \ref{tab:fit2}.  In all
cases there is at least $6.1 \sigma$ evidence for multipoles beyond $E1$
in the transition $\psip \to \gamma' \chi_{c2} \to \gamma' \gamma J/\psi$.
The contours for $\sqrt{2 \Delta \log {\cal L}}$ for $a_2$ vs $b_2$ for the 
two-parameter fits are shown in Fig.~\ref{fig:contour}(b); the
contours for all other pairs of variables for all the fits are Gaussian-shaped
with a single local maximum.

\begin{table}[htb]
\caption{$J_{\chi}=2$ five-angle fit results.  The fits were performed on the
19755 signal events satisfying the selection criteria described in Sec.\
\ref{selection_criteria}. $\chi_{E1} \equiv \sqrt{2\Delta \log
\mathcal{L}}$ is the number of standard deviations by which the fitted value
differs from the pure $E1$ value.
\label{tab:fit2}}
\begin{center}
\begin{tabular}{lccccc} \hline \hline
  \rule[12pt]{-1mm}{0mm}
Fit & $a_2^{J=2}$ & $b_2^{J=2}$ & $a_3^{J=2}$ & $b_3^{J=2}$ & $\chi_{E1}$ \\
 & $(10^{-2})$ & $(10^{-2})$ & $(10^{-2})$ & $(10^{-2})$ & \\ \hline
Two-parameter   & $-9.3 \pm 1.6$ & $1.0 \pm 1.3$ & 0  & 0  & 6.2 \\
Three-parameter & $-9.3 \pm 1.6$ & $0.7 \pm 1.3$ & 0  & $-0.8 \pm 1.2$ & 6.3 \\
Two-param.\ ($b_2 = \frac{-a_2}{3.367}$) & $-9.2 \pm 1.6$ & $2.7 \pm 0.5$ & 0 
 & $-0.1 \pm 1.1$ & 6.1 \\
Four-parameter & $-7.9 \pm 1.9$ & $0.2 \pm 1.4$ & $1.7 \pm 1.4$ & $-0.8 \pm 1.2$ & 6.4 \\
Theory ($m_c = 1.5 \gev$)& {~~$-9.6(1 + \kappa_c)$~~}
 & {~~$2.9(1 + \kappa_c)$~~} & 0 & {~~Model dep.~~} & \\ \hline \hline
\end{tabular}
\end{center}
\end{table}

\section{SYSTEMATIC UNCERTAINTIES \label{sec:sys}}

We now present the results of systematic studies for the fits to the five-angle
distributions performed in the previous section.  For $J_\chi = 1$, we perform
all systematic studies on the two-parameter fit ($a_2, b_2$), as the
one-parameter fixed-ratio fit produces nearly identical results.  However, for
$J_\chi = 2$, there are four types of five-angle fits:

\begin{itemize}
\item Two-parameter fit ($a_2, b_2$) with $a_3 \equiv b_3 \equiv 0$ (no S--D
or P--F mixing),
\item Three-parameter fit ($a_2, b_2, b_3)$ with $a_3 \equiv 0$, relevant for a
D-wave admixture in the $\psip$,
\item Fixed-ratio two-parameter fit ($a_2, b_3$) with $b_2 \equiv -a_2/3.367$
and $a_3 \equiv 0$, and
\item Four-parameter fit ($a_2, b_2, a_3, b_3$).
\end{itemize}
In this paper, we describe in detail the systematic studies for the 
$J_\chi = 1$ case and the $J_\chi = 2$ case where $a_3 = b_3 \equiv 0$.  
Systematic studies for the other three $J_\chi = 2$ cases are discussed in 
detail in Ref.\ \cite{E1M2E3_thesis}.

For many investigations into a possible systematic uncertainty, we perform an 
ensemble of fits on samples of signal events selected from a phase space data
set via the rejection method to follow $\PDFz$ for a given input set of
multipole parameters.  For each multipole $a$, we calculate the following
parameters from the results of these ensembles of fits, with $N\tsb{ens}$ MC 
events in each member of the ensemble: 
\begin{itemize}
 \item $\langle a \rangle$, the mean of the fitted multipole amplitude over
the ensemble of tests, with a statistical error corresponding to the
variation of the fitted multipole amplitude over the ensemble of tests,
 \item $\sigma\tsp{fit}_{a}$, the (mean of the) nominal uncertainty from each
individual likelihood fit to multipole amplitude,
 \item $\Delta_{\langle a \rangle}$, the deviation of the mean from the
MC-generated value of the amplitude in units of the expected deviation of the
mean $\sigma_{\langle a \rangle} = \sigma\tsp{fit}/\sqrt{N\tsb{ens}}$ defined
as:
\beq
 \Delta_{\langle a \rangle} = \frac{\langle a \rangle - a\tsp{Input}}{
\sigma_a\tsp{fit} / \sqrt{N\tsb{ens}}}~\text{, and}
\eeq
\item  $\Delta_{\sigma(a)}$, the deviation of the standard deviation when a
potential systematic effect is present compared to the standard deviation 
without the effect being present (in units of the expected fluctuation the 
best estimate of the standard deviation from an ensemble of $N$ measurements 
$\sigma_{\sigma} = \sigma/\sqrt{2N}$ (\cite{PDG} Sec.\ 32.1.1), defined as: 
\beq
 \Delta_{\sigma(a)} = \frac{\sigma\tsp{with\;syst}
 - \sigma\tsp{without\;syst}}{\sigma / \sqrt{2 N\tsb{ens}}} \quad \text{.}
\eeq
\end{itemize}
For $\sigma\tsp{fit}$ we list the mean of the nominal uncertainty, but
for all tests performed the nominal uncertainty from every likelihood fit 
in the ensemble was essentially constant to the level of precision quoted.

For all of the systematic tests from ensembles of measurements, we assign a 
systematic uncertainty if either (a) we find that there is a significant bias 
$|\Delta_{\langle a\rangle}| > 1$ or if (b) there is an uncertainty that 
widens the ensemble distribution above the expected statistical fluctuation 
evidenced by $\Delta_{\sigma(a)} > 1$.

\subsection{Toy MC check of fitting procedure}

To test the accuracy of the fitting procedure described in Sec.\
\ref{fitting_procedure} an ensemble of toy Monte Carlo fitting trials was
performed.  For each trial, we generated a large number of phase space
events, where each event is described by five random numbers for each of the
variables ($\fivevars$) uniformly distributed over their ranges.  

We generated a set of toy signal Monte Carlo events by selecting events from a
separate MC phase space data set via the rejection method,
so the events are described by $\PDFz$ for an input set of
multipole parameters $\pmbA_0$.  

To test the $J_{\chi} = 1$ ($J_{\chi} = 2$) fits, we performed an ensemble of
200 toy MC trials in which each trial had $N\tsb{sig} = 40000$ (20000)
signal events after selection criteria were applied. We analytically calculated
the phase space integrals, as the toy MC was thrown at 100\% detector
efficiency.
In assigning the systematic uncertainty, we set the multipole amplitudes of the 
toy signal Monte Carlo to be $a_2^{J=1} = -0.065$, $b_2^{J=1} = 0.029$, 
($a_2^{J=2} = -0.096$, $b_2^{J=2} = 0.029$, $a_3^{J=2} = 0$, $b_3^{J=2} = 0$)
the expected value if $\kappa_c = 0$ to first order in $E_\gamma/m_c$.  
Fits with other values of input parameters recover the input results to 
similar precision.  We find no systematic bias or uncertainty is associated
with the fitting procedure described in this method, as the ensemble of trials
is Gaussian-distributed with a width according to the statistical uncertainty.

\subsection{Amount of phase space Monte Carlo needed for efficiency integrals}

Using too few phase space Monte Carlo events would give poor approximations to
the efficiency integrals, introducing an overall systematic uncertainty to the
results of the maximum likelihood fit.  We use 4.5 million phase space events
for the normalization, approximately 100 times the $J_\chi = 1$ data set (the
larger of the two).  By varying the size of the MC sample, we find no
systematic uncertainty associated with any number of events exceeding $10^5$,
and hence assign no systematic error to this source.

\subsection{Impurity systematic uncertainties}

For the $J_\chi = 1$ ($J_\chi = 2$) selection criteria, approximately $0.23\%$
($0.29\%$) of the events that pass the selection criteria are not signal
events, but a background mode that must be considered for the possibility of
introducing a systematic bias or uncertainty to our result.  Taking our
five-fold generic Monte Carlo data set and splitting it into five independent
data sets, we find
a purity and efficiency of 99.77\% and 39.6\% (99.71\% and 36.0\%).  The main
sources of impurity background modes for $J_\chi = 1$ are $\psip \to \pi^0
\pi^0 \jpsi$ and $\psip \to \gamp \chic{1}$ (where the $\chic{1}$ decay was not
to $\gamma \jpsi$ followed by $J/\psi \to \ell^+ \ell^-$).  For $J_\chi = 2$
they are $\psip \to \gamp \chic{1}$ and $\psip \to \pi^0 \pi^0 \jpsi$.

For each of the five independent generic MC impurity backgrounds, we perform
31 (37) trials with and without the impurity background events
present.  For each trial, we replace the signal events originally present with
phase space events selected via the rejection method to come up with many
independent data sets.  For each trial we perform one fit with no impurities
present and one fit with the impurities.  For a given set of impure
background events, we find that the bias due to impurities varies very little 
between different trials.  In Table \ref{j1_genmc_impurity_table}
(Table \ref{j2_genmc_impurity_table_a2_b2}), we list the difference from the
fit with no impurities.  For $J_\chi = 1$, we find a significant impurity bias
that is relatively constant among all five sets of impure events, so we correct
our fitted result for this impurity bias and assign a systematic uncertainty of
half of the bias.  For the $J_\chi = 2$ case, we find that the impurity bias
significantly fluctuates between background data sets, so we assign a
systematic uncertainty of the size of the fluctuation of the impurity bias.

\begin{table}[htb]
 \caption{Generic MC tests for a systematic bias from impure events
for $J_\chi = 1$. 
We split 
the five-fold generic MC data set into five data sets labeled (A)-(E), replacing
the generic Monte Carlo signal events with events selected to obey $\PDFz$ from the 
4.5M event phase space MC data set.  For each of
these five data sets, we performed an ensemble of thirty-one fits.  
The difference rows show the shift in values of $a_2$ and $b_2$ 
comparing the individual fits
before and after impurities are added.  A positive shift means that to obtain
the pure results we should subtract the bias from impurities.  The set
(A-E) is the result from adding all five data samples of impure events to a
regular-sized set of signal events, and demonstrates how the impurities scale
linearly in the $J_\chi = 1$ case.}
\label{j1_genmc_impurity_table}
\begin{center}
\begin{tabular}{lcccccc} \hline \hline
  \rule[12pt]{-1mm}{0mm}
Type & $\langle a_2 \rangle$ & 
~$\langle \sigma_{a_2} \tsp{fit} \rangle$~ & ~$\Delta_{\langle a_2 \rangle}$~ 
& $\langle b_2 \rangle$ & 
~$\langle \sigma_{b_2} \tsp{fit} \rangle$~ & ~$\Delta_{\langle b_2 \rangle}$~ \\
 & $(10^{-2})$ & $(10^{-2})$ &  & $(10^{-2})$ & $(10^{-2})$ & \\ \hline
Pure
 & $-6.54 \pm 0.50$~~ & 0.63 & -0.32 & 
   $ 2.97 \pm 0.71$ & 0.73 & 0.52 \\
Difference w/impurities (A)
 & $0.150 \pm 0.002$ &  &  &
   $0.058 \pm 0.003$ &  &  \\ 
Difference w/impurities (B) 
 & $0.120 \pm 0.002$ &  &  &
   $0.053 \pm 0.003$ &  &  \\ 
Difference w/impurities (C)
 & $0.140 \pm 0.003$ &  &  &
   $0.060 \pm 0.005$ &  &  \\ 
Difference w/impurities (D)
 & $0.216 \pm 0.004$ &  &  &
   $0.095 \pm 0.005$ &  &  \\ 
Difference w/impurities (E) 
 & $0.109 \pm 0.002$ &  &  &
  $-0.031 \pm 0.003$~~ &  &  \\
Difference w/impurities (A-E)
 & $0.730 \pm 0.011$ &  &  &
   $0.241 \pm 0.019$ &  &  \\ 
  Input    & $-6.50$~~~~~~~~~~~ & & & $2.90$~~~~~~~~~ & &  \\ 
$\langle \text{Impurity bias} \rangle$ & $0.15 \pm 0.03$ & & & $0.05 \pm 0.03$ & & \\
\hline \hline
 \end{tabular} 
\end{center}
\end{table}

\begin{table}[htb]
 \caption{Generic MC tests for a systematic bias from impure events for
$J_\chi = 2$ for two-parameter ($a_2, b_2$) fit with $a_3\equiv b_3 \equiv 0$.
We find that the impurities add a negligible systematic uncertainty when
compared with the statistical uncertainty.}
\label{j2_genmc_impurity_table_a2_b2}
\begin{center}
\begin{tabular}{lcccccc} \hline \hline
  \rule[12pt]{-1mm}{0mm}
Type& $\langle a_2 \rangle$ & 
~$\langle \sigma_{a_2} \tsp{fit} \rangle$~ & ~$\Delta_{\langle a_2 \rangle}$~ & 
$\langle b_2 \rangle$ & 
~$\langle \sigma_{b_2} \tsp{fit} \rangle$~ & ~$\Delta_{\langle b_2 \rangle}$~ \\
 & $(10^{-2})$ & $(10^{-2})$ &  & $(10^{-2})$
 & $(10^{-2})$ & \\ \hline
\hline Pure 
 & $-9.8 \pm 1.4$~~ & 1.6 & -0.6 & 
   $ 3.0 \pm 1.3$ & 1.2 & 0.6 \\
Difference w/impurities (A)
 & $-0.005 \pm 0.006$~~ &  &  &
    $0.078 \pm 0.003$ &  &  \\ 
Difference w/impurities (B) 
 & $0.080 \pm 0.004$ &  &  &
  $-0.011 \pm 0.005$~~ &  &  \\ 
Difference w/impurities (C) 
 & $-0.008 \pm 0.011$~~ &  &  &
    $0.149 \pm 0.004$ &  &  \\ 
Difference w/impurities (D)
 & $0.022 \pm 0.003$ &  &  &
  $-0.050 \pm 0.003$~~ &  &  \\ 
Difference w/impurities (E) 
 & $-0.041 \pm 0.002$~~ &  &  &
    $0.027 \pm 0.003$ &  &  \\ 
Difference w/impurities (A-E) 
 & $0.047 \pm 0.019$ &  &  &
   $0.190 \pm 0.011$ &  &  \\ 
Input    & $-9.6$~~~~~~~~~~  & & & $2.9$~~~~~~~~  & &  \\
$\langle \text{Impurity Bias} \rangle$ & $0.009 \pm 0.040$ & & & $0.038 \pm 0.070$ & & \\
\hline \hline
\end{tabular}
\end{center}
\end{table}

\subsection{Final state radiation}
Another possible source of systematic uncertainty is the effect of final state
radiation (FSR), which can alter the directions of the two leptons in
the $\jpsi$ rest frame affecting the variables $\costh$ and $\phi$.
Generation of Monte Carlo samples has been done using EvtGen, which models
final state radiation in the decay sequences $\jpsi \to \ell^+ \ell^-$ with
PHOTOS.  We estimate the effect of final state radiation by performing signal
fits on the angles $\Omega$ from generator level four-vectors, both before and
after final state radiation has been added.  We use the rejection method
to select events, so that the pre-FSR generator level
four-vectors follow the PDF $\PDFz$ for an input value of the multipole
amplitudes $\pmbA_0$.  We also use the pre-FSR four-vectors when selecting the
phase space events to be used as `signal' described by the PDF $\PDFz$ with a
given $\pmbA_0 \equiv (a_2, b_2) = (-0.065, 0.029)$ (for $J_\chi = 2$, $\pmbA_0
\equiv (a_2, b_2, a_3, b_3) = (-0.096, 0.029, 0.0, 0.0)$).  We then compare the
fit on the selected events using the pre-FSR and post-FSR generator level to
check for a systematic uncertainty from final state radiation.  Comparing 
$\Delta_{\sigma(a)}$ and $\Delta_{\langle a \rangle}$ for each multipole
parameter for FSR, we find no statistically significant evidence for a
systematic uncertainty due to Final State Radiation.

The angular distribution was also fit using data from $\jpsi \to e^+ e^-$ and 
$\jpsi \to \mu^+ \mu^-$ only, selected by the $E/p$ selection criterion.  
Without correcting for possible systematic biases, we found the results in 
Table \ref{table:muon_e_only}.  Preliminary studies of simulated signal data 
(generated from phase space MC) indicated that the most accurate result is
obtained by performing a fit to the combined $\jpsi \to \mu^+ \mu^-$ and 
$\jpsi \to e^+ e^-$ dataset, while still blinded to the actual dataset.  The 
results from the fit to the muon-only dataset were similar to the results 
from the combined dataset.  The electron-only dataset produces similar 
results to the fit results of the combined dataset, except for 
the $\chi_{c1}$ case where the two-parameter electron-only result deviates
from the combined result by approximately 1.4$\sigma$.  However, fixing the 
$a_2/b_2$ ratio reduces the deviation of the electron-only result to less than
1$\sigma$ even in this worst case.  Therefore, we assign no additional 
systematic uncertainty due to FSR from the results of the muon-only and 
electron-only fits.

\begin{table}
\caption{Fits to the angular distribution using only events where the $\jpsi$
decays to two muons or two electrons. The $\chi_{c1}$ ($\chi_{c2}$) 
dataset contained 20968 (10563) muon-only events and 18395 (9192) 
electron-only events.}
\label{table:muon_e_only}
\begin{center}
\begin{tabular}{lcccc}
\hline \hline
Fit & $a_2$ & $b_2$ & $a_3$ & $b_3$ \\
& $(10^{-2})$ & $(10^{-2})$ & $(10^{-2})$ & $(10^{-2})$ \\
\hline
$\chi_{c1}$ Two-param $\mu \mu$ & $-7.23 \pm 0.87$ & $2.2 \pm 1.0$ & & \\
$\chi_{c1}$ Fixed-ratio $\mu \mu$ & $-6.85 \pm 0.75$ & $3.0 \pm 0.3$ & & \\
$\chi_{c1}$ Two-param $e e$ & $-4.85 \pm 0.92$ & $3.5 \pm 1.0$ & & \\
$\chi_{c1}$ Fixed-ratio $e e$ & $-5.36 \pm 0.81$ & $2.4 \pm 0.4$ & & \\ \hline
$\chi_{c2}$ Two-param $\mu \mu$ & $-8.1 \pm 2.1$ & $1.2 \pm 1.7$ &0 &0 \\
$\chi_{c2}$ Three-param $\mu \mu$ & $-8.1 \pm 2.1$ & $1.1 \pm 1.9$ 
&0 & $-0.3\pm 1.6$ \\
$\chi_{c2}$ Fixed-ratio $\mu \mu$ & $-8.1 \pm 2.1$ & $2.4 \pm 0.6$ 
&0 & ~~~$0.2 \pm 1.5$ \\
$\chi_{c2}$ Four-param $\mu \mu$ & $-5.4 \pm 3.0$ & $0.0 \pm 2.0$ 
& $3.6 \pm 1.8$ & $-0.2 \pm 1.7$ \\
$\chi_{c2}$ Two-param $e e$ & $-10.7 \pm 2.3$ & $0.8 \pm 1.8$ &0 &0 \\
$\chi_{c2}$ Three-param $e e$ & $-10.7 \pm 2.3$ & $0.2 \pm 2.0$ 
&0 & $-1.4\pm 1.8$ \\
$\chi_{c2}$ Fixed-ratio $e e$ & $-10.5 \pm 2.3$ & $3.1 \pm 0.7$ 
&0 & $-0.4 \pm 1.6$ \\
$\chi_{c2}$ Four-param $e e$ & ~$-11.2 \pm 3.0$~ & ~$0.4 \pm 2.1$~ 
& ~$-0.6 \pm 2.2$~ & ~$-1.4 \pm 1.8$~ \\
\hline \hline
\end{tabular}
\end{center}
\end{table}
 
\subsection{Choice of kinematic fits}

For our final analysis, we perform a 1C kinematic fit to the $\jpsi$ mass and a
4C kinematic fit to the $\psip$ four momentum of the lab frame, and also
perform bremsstrahlung reconstruction on each track if any showers were tagged
as bremsstrahlung radiation  belonging to the track.  To test for possible
systematic effects, we perform an ensemble of tests on phase space MC shaped to
have $\pmbA_0 = (-0.065, 0.029)$ for $J_\chi =1$ and $\pmbA_0 = (-0.096, 0.029,
0, 0)$ for $J_\chi = 2$ with four-vectors selected to have the pre-FSR
generator photons follow $\PDFz$.  We construct the four-vectors for the
variables in three ways: (1) Post-FSR generator level four-vectors;
(2) 1C and 4C kinematic fits without bremsstrahlung recovery; (3) 1C and 4C
kinematic fits with bremsstrahlung recovery.  For each four-vector type, we
perform as many fits as possible using a data size (after selection criteria)
of 40000 $J_\chi\!=\!1$ (20000 $J_\chi\!=\!2$) events in each fit.  We find no
statistically significant systematic effect from this procedure.

\subsection{Variation of selection criteria}

To look for an additional systematic uncertainty from possible variations of
selection criteria, we looked at effects of the following variations on
statistical and systematic impurity uncertainties: maximum third shower
energy, maximum reduced $\chi^2$, $\chi_c$ mass window, and maximum cosine
of polar angle for photons in the barrel region.  Variations were explored
which loosened and tightened all our selection criteria.
For $J_\chi = 1$ we found that the default criteria (defined in Sec.\
\ref{selection_criteria}) had the smallest quadrature sum of the statistical
uncertainty with impurity systematic uncertainty.  We further found that over
the ensemble of tests involving various criteria, the mean from the ensemble of
tests for $a_2$ and $b_2$ (when no impurities were present) varied only
slightly.  For the $J_\chi = 2$ two-parameter ($a_2, b_2$) fit case, we found
that while we were quite near the minimal total quadrature sum for the default
criteria, we could have achieved a $\sim 3\%$ improvement if we loosened these
conditions.  However, to achieve that $\sim 3\%$ improvement requires
increasing the number of impure events by a factor of approximately five as
shown in Fig. \ref{fig:maxredchisq}, so this was not performed.

After looking at the effect of variations of selection criteria on an ensemble
of tests using the ``signal'' data selected from phase space MC via the
rejection method, we looked at the actual effect of performing fits to data
after applying various criteria.  These results show the sensitivity of the
data to the chosen criteria.  For the $J_\chi = 1$ case shown in Table
\ref{j1_data_cut_var}, we perform the fits using the various criteria, and then
correct for the impurity bias.  We then consider the ensemble of bias-corrected
data fits aed assign a systematic uncertainty using the standard deviation of
the fitted results over the 7 types of criteria considered.  We find a
systematic uncertainty of $(0.19, 0.22)\pow{-2}$ for ($a_2^{J=1}, b_2^{J=1}$)
in performing fits to data.

\begin{table}[htb]
\caption{Results of data fits for $J_\chi =1$ when applying various selection
criteria.  For all selection criteria considered, a systematic uncertainty is
found of $(0.19, 0.22) \pow{-2}$ for $(a_2, b_2)$, respectively, over the
variation of the criteria.}
\label{j1_data_cut_var}
\begin{center}
\begin{tabular}{lccc} \hline \hline
  \rule[12pt]{-1mm}{0mm}
Criteria & $a_2\tsp{bias\, cor}$ & & $b_2\tsp{bias\, cor}$ \\
& $(10^{-2})$ & & $(10^{-2})$ 
\\ \hline
Default  & 
~$-6.26 \pm 0.63 \pm 0.15$~  & & ~$2.76 \pm 0.73 \pm 0.06$~  \\
$E\tsp{3rd\; Shwr} < 18\mev$  & 
$-6.43 \pm 0.64 \pm 0.08$   & & $2.67 \pm 0.73 \pm 0.06$  \\
$E\tsp{3rd\; Shwr} < 50\mev$  & 
$-5.73 \pm 0.60 \pm 0.30$   & & $2.45 \pm 0.72 \pm 0.13$  \\
$\chi_{k.f.}^2 < 10$ & 
$-6.23 \pm 0.65 \pm 0.05$    & & $2.33 \pm 0.75 \pm 0.03 $            \\
$\chi_{k.f.}^2 < 30$  & 
$-6.30 \pm 0.61 \pm 0.32$ & & $3.10 \pm 0.71 \pm 0.04$  \\
$\chi_c$ mass $\pm 10 \mev$ & 
$-6.36 \pm 0.65 \pm 0.11$  & & $2.85 \pm 0.75 \pm 0.04$ \\
$\chi_c$ mass $\pm 20 \mev$ & 
$-6.10 \pm 0.62 \pm 0.18$  & & $2.78 \pm 0.69 \pm 0.09$ \\
$|\costh\tsp{barrel}\tsb{lab, ph}| < 0.77$ & 
$-6.18 \pm 0.65 \pm 0.16$ & & $2.97 \pm 0.75 \pm 0.07$ \\
$|\costh\tsp{barrel}\tsb{lab, ph}| < 0.80$ & 
$-6.17 \pm 0.62 \pm 0.16$ & & $2.73 \pm 0.72 \pm 0.07$ \\
Ensemble & $-6.20 \pm 0.19 $ & & $2.74 \pm 0.22$ \\ \hline \hline
\end{tabular}
\end{center}
\end{table}

For $J_\chi = 2$ (Table \ref{j2_data_cut_var_a2_b2}), we follow a similar
procedure but do not
correct for impurity biases before calculating the systematic uncertainty, as
the impurity bias in all cases is less than 1/10 the statistical uncertainty,
so any correction would be of very little significance.  We find in this case
systematic uncertainties of $(0.3, 0.3) \pow{-2}$ for ($a_2, b_2)$ when
performing the two-parameter fit with ($a_3\equiv b_3 \equiv 0$).

\begin{table}[htb]
\caption{Results of data fits when applying various selection criteria to
$J_\chi=2$ two-parameter ($a_2, b_2$) fits ($a_3 \equiv b_3 \equiv 0$).  For
all sets of criteria, a systematic uncertainty is found of $(0.3, 0.3)\pow{-2}$
for $(a_2, b_2)$, respectively, over the variation of the criteria.}
\label{j2_data_cut_var_a2_b2}
\begin{center}
\begin{tabular}{lccc} \hline \hline
Criteria & $a_2$ & & $b_2$ \\
& $(10^{-2})$ & & $(10^{-2})$ \\ \hline
Default & ~$-9.3 \pm 1.6$~ & & ~$1.0 \pm 1.3$~ \\ 
$E\tsp{3rd\; Shwr} < 18\mev$ 
 & $-9.4 \pm 1.6$ & & $0.6 \pm 1.3$ \\ 
$E\tsp{3rd\; Shwr} < 50\mev$ 
 & $-9.8 \pm 1.6$ & & $0.5 \pm 1.3$ \\ 
$\chi\tsb{k.f.}^2 < 10$  
 & $-9.1 \pm 1.6$ & & $1.3 \pm 1.3$ \\ 
$\chi\tsb{k.f.}^2 < 30$ 
 & $-9.5 \pm 1.5$ & & $0.4 \pm 1.2$ \\ 
$\chi_c$ mass $\pm 10 \mev$ 
 & $-8.7 \pm 1.6$ & & $1.0 \pm 1.3$ \\ 
$\chi_c$ mass $\pm 20 \mev$
 & $-9.8 \pm 1.5$ & & $0.8 \pm 1.3$ \\ 
$|\costh\tsp{barrel}\tsb{lab, ph}| < 0.77$ 
 & $-9.6 \pm 1.6$ & & $1.2 \pm 1.3$ \\ 
$|\costh\tsp{barrel}\tsb{lab, ph}| < 0.80$ 
 & $-9.5 \pm 1.5$ & & $1.3 \pm 1.3$ \\ 
Ensemble & $-9.4 \pm 0.3$ & & $0.9 \pm 0.3$ \\ \hline \hline
\end{tabular}
\end{center}
\end{table}

\subsection{Summary of systematic uncertainties and biases}

The systematic uncertainties and biases for $J_\chi = 1$ are summarized in
Table \ref{j1_systematic_table}.  We find the total systematic uncertainty to
be $(0.24, 0.23)\pow{-2}$ for $(a_2^{J=1}, b_2^{J=1})$, respectively.
The systematic uncertainties for the $J_\chi = 2$ two-parameter fit
($a_2, b_2$) are summarized in Table \ref{j2_systematic_table_a2_b2}, and
for other fits in Tables \ref{j2_systematic_table_a2_b2_b3}--%
\ref{j2_systematic_table_a2_b2_a3_b3}.  In each case the total
systematic error is the quadrature sum of systematic uncertainties, and the
statistical uncertainty for the data fits is given for comparison.  We do
not find any systematic biases for the $J_\chi = 2$ case.

\begin{table}[htb]
\caption{Systematic uncertainties and biases for $J_\chi = 1$.  The total
systematic error is the quadrature sum of systematic uncertainties and the
signed sum of systematic biases.  The statistical uncertainty from the data
fits is given for comparison.}
\label{j1_systematic_table}
\begin{center}
\begin{tabular}{lcccc} \hline \hline
  \rule[12pt]{-1mm}{0mm}
Systematic uncertainty & \multicolumn{2}{c}{$a_2^{J = 1}$}
 & \multicolumn{2}{c}{$b_2^{J = 1}$} \\
 & Uncertainty & Bias & Uncertainty & Bias \\
 & $(10^{-2})$ & $(10^{-2})$ & $(10^{-2})$ & $(10^{-2})$ \\ \hline
Generic MC impurities & 0.15 & 0.15 & 0.05 & 0.05 \\
Selection criteria & 0.19 & - & 0.22 & - \\
Total systematic uncert.\ & 0.24 & 0.15 & 0.23 & 0.05 \\
Statistical uncertainty & 0.63 & - & 0.73 & - \\ \hline \hline
\end{tabular}
\end{center}
\end{table}

\begin{table}[htb]
\caption{Systematic uncertainties for $J_\chi = 2$ two-parameter fit with
$a_2, b_2$.}
\label{j2_systematic_table_a2_b2}
\begin{center}
\begin{tabular}{lcc} \hline \hline
  \rule[12pt]{-1mm}{0mm}
Systematic uncertainty & $a_2^{J=2}$ & $b_2^{J=2}$ \\
  & ~$(10^{-2})$~ & ~$(10^{-2})$~ \\ \hline
Generic MC impurities & 0.04 & 0.07 \\
Selection criteria & 0.33 & 0.33 \\
Total systematic uncert.\ & 0.3 & 0.3 \\
Statistical uncertainty & 1.6 & 1.3 \\ \hline \hline
\end{tabular}
\end{center}
\end{table}

\begin{table}[htb]
\caption{Systematic uncertainties for $J_\chi = 2$ three-parameter fit for
$a_2, b_2, b_3$.}
\label{j2_systematic_table_a2_b2_b3}
\begin{center}
\begin{tabular}{lccc} \hline \hline
  \rule[12pt]{-1mm}{0mm}
Systematic uncertainty & $a_2^{J=2}$ & $b_2^{J=2}$ & $b_3^{J=2}$ \\
  & ~$(10^{-2})$~ & ~$(10^{-2})$~ & ~$(10^{-2})$~ \\ \hline
Generic MC impurities & 0.04 & 0.07 & 0.03 \\
Selection criteria & 0.33 & 0.34 & 0.20 \\
Total systematic uncert.\ & 0.3 & 0.3 & 0.2 \\
Statistical uncertainty & 1.6 & 1.4 & 1.2 \\ \hline \hline
\end{tabular}
\end{center}
\end{table}
\clearpage

\begin{table}[htb]
\caption{Systematic uncertainties for $J_\chi = 2$ two-parameter fit for
$a_2, b_3$ with fixed values of $b_2 \equiv -a_2/3.367$ and $a_3 \equiv 0$.}
\label{j2_systematic_table_a2_b3}
\begin{center}
\begin{tabular}{lcc} \hline \hline
  \rule[12pt]{-1mm}{0mm}
Systematic uncertainty & $a_2^{J=2}$ & $b_3^{J=2}$ \\
  & ~$(10^{-2})$~ & ~$(10^{-2})$~ \\ \hline
Generic MC impurities & 0.04 & 0.04 \\
Selection criteria & 0.34 & 0.23 \\
Total systematic uncert.\ & 0.3 & 0.2 \\
Statistical uncertainty & 1.6 & 1.1 \\ \hline \hline
\end{tabular}
\end{center}
\end{table}


\begin{table}[htb]
\caption{Systematic uncertainties for $J_\chi = 2$ four-parameter fit with
$a_2, b_2, a_3, b_3$.}
\label{j2_systematic_table_a2_b2_a3_b3}
\begin{center}
\begin{tabular}{lcccc} \hline \hline
  \rule[12pt]{-1mm}{0mm}
Systematic uncertainty & $a_2^{J=2}$ & $b_2^{J=2}$ & $a_3^{J=2}$
 & $b_3^{J=2}$ \\
 & $(10^{-2})$ & $(10^{-2})$  & $(10^{-2})$ & $(10^{-2})$ \\ \hline
Generic MC impurities & 0.06 & 0.08 & 0.08 & 0.03 \\
Selection criteria & 0.24 & 0.39 & 0.28 & 0.20 \\
Total systematic uncert.\ & 0.3 & 0.4 & 0.3 & 0.2 \\
Statistical uncertainty & 1.9 & 1.5 & 1.4 & 1.2 \\ \hline \hline
\end{tabular}
\end{center}
\end{table}

\section{CONCLUSIONS \label{sec:concl}}

\subsection{Normalized multipole amplitudes}

The results of our bias-corrected fits with systematic uncertainties for
$J_\chi = 1$ with the two-parameter fit are:
\bea
a_2^{J=1} &=& (-6.26 \pm 0.63 \pm 0.24) \pow{-2}~, \\
b_2^{J=1} &=& (2.76 \pm 0.73 \pm 0.23) \pow{-2}~.
\eea

The results of our fits with systematic uncertainties for $J_\chi = 2$ with the
two-parameter fit ($a_2, b_2$) with $a_3 = b_3 \equiv 0$ are:
\bea
 a_2^{J=2} &=& (-9.3 \pm 1.6 \pm 0.3) \pow{-2}~, \\
 b_2^{J=2} &=& ( 1.0 \pm 1.3 \pm 0.3) \pow{-2}~;
\eea
for the three-parameter fit ($a_2,b_2,b_3$) with $a_3 \equiv 0$:
\bea
 a_2^{J=2} &=& (-9.3 \pm 1.6 \pm 0.3) \pow{-2}~, \\
 b_2^{J=2} &=& ( 0.7 \pm 1.4 \pm 0.3) \pow{-2}~, \\
 b_3^{J=2} &=& (-0.8 \pm 1.2 \pm 0.2) \pow{-2}~;
\eea
for the two-parameter fit ($a_2, b_3$) with fixed values of 
$b_2 \equiv -a_2/3.367$ and $a_3 \equiv 0$:
\bea
  a_2^{J=2} &=& (-9.2 \pm 1.6 \pm 0.3) \pow{-2}~, \\
 b_2^{J=2} &\equiv& -\frac{a_2^{J=2}}{3.367} = ( 2.7 \pm 0.5 \pm 0.1) 
\pow{-2}~, \\
 b_3^{J=2} &=& (-0.1 \pm 1.1 \pm 0.2)\pow{-2}~;
\eea
and for the four-parameter fit ($a_2,b_2,a_3,b_3$):
\bea
 a_2^{J=2} &=& (-7.9 \pm 1.9 \pm 0.3) \pow{-2}~,\\
 b_2^{J=2} &=& ( 0.2 \pm 1.5 \pm 0.4) \pow{-2}~, \\
 a_3^{J=2} &=& ( 1.7 \pm 1.4 \pm 0.3) \pow{-2}~, \\
 b_3^{J=2} &=& (-0.8 \pm 1.2 \pm 0.2)\pow{-2}~.
\eea

\begin{figure}[htb]
\begin{center}
\includegraphics[width=5.25in]{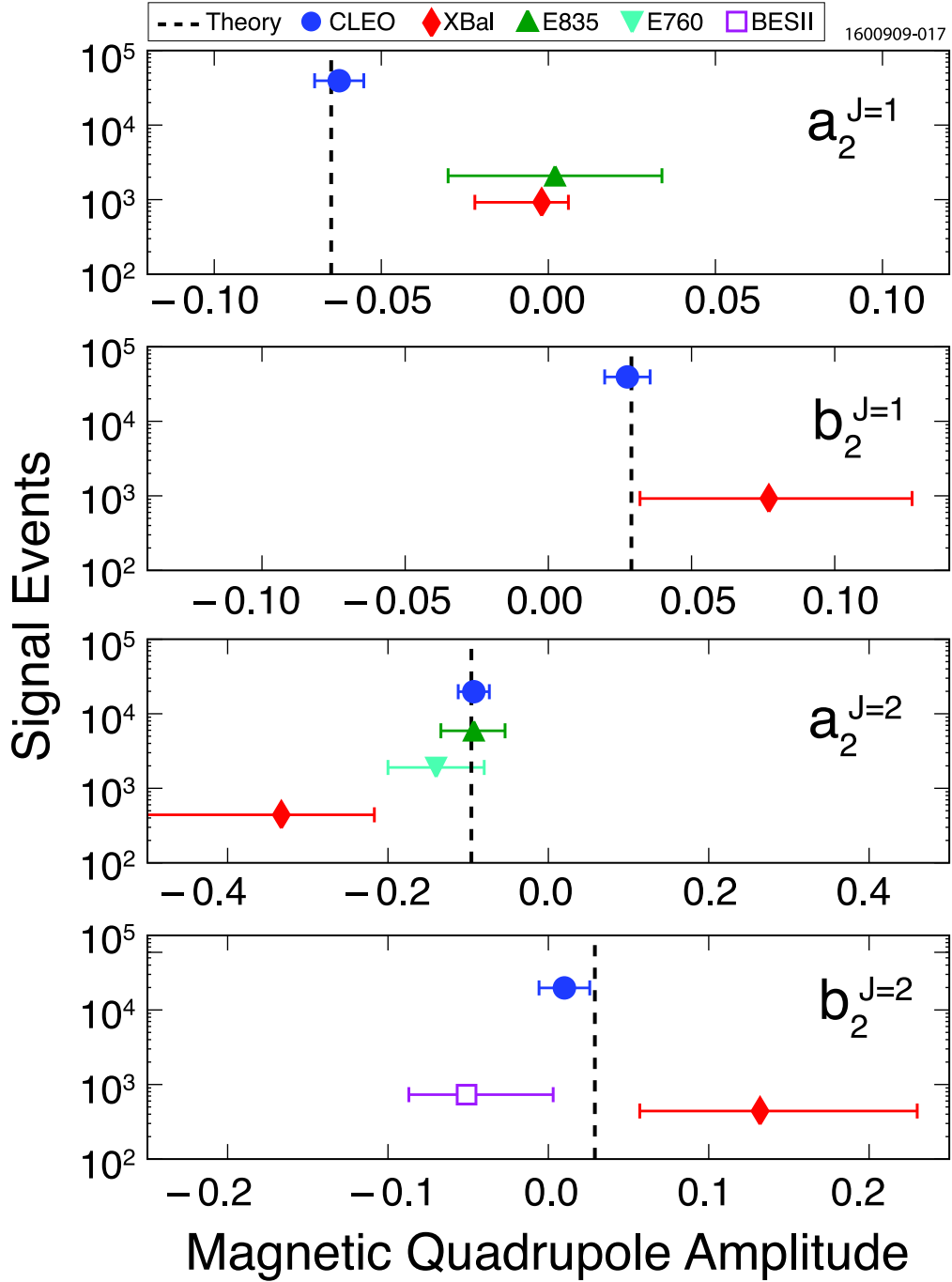}
\end{center}
\caption{Experimental values of the magnetic quadrupole amplitudes from this
analysis compared with previous experimental values and theoretical
expectations.  For $J_\chi=2$, results are shown for the two-parameter fit.
CLEO-c results from this analysis are solid circles; Crystal Ball results are
diamonds \cite{Oreglia:1982}, the E760 result is a $\bigtriangledown$ 
\cite{E760:1993}, the E835 results are $\bigtriangleup$'s \cite{E835:2002}, 
the BESII result is an open
square \cite{BES:2004}, and the theoretical expectations given by Eqs.\
(\ref{eq:a2J1})--(\ref{eq:b2J2}) with $m_c = 1.5\gev$ and $\kappa_c = 0$ are
dashed lines.}
\label{fig:all_expt_results}
\end{figure}

Our results are compared with previous experiments and theory in Fig.\
\ref{fig:all_expt_results}.  The $J_\chi=2$ results shown are for the
two-parameter fit with $a_3 = b_3 = 0$.


\subsection{\boldmath Ratios independent of $m_c$ and $\kappa_c$}

Using the results from the $J_\chi = 2$ two-parameter ($a_2, b_2$) fit and the
$J_\chi = 1$ fit, we find the ratios with the highest statistical sensitivity 
compare very well with the theoretical predictions:
\bea
\left(\frac{a_2^{J=1}}{a_2^{J=2}} \right)\tsb{CLEO} = ~~ 0.67^{+0.19}_{-0.13} 
&\stackrel{?}{=}&  \left(\frac{a_2^{J=1}}{a_2^{J=2}} \right)\tsb{th} = 
0.676 \pm 0.071~, \\
\left(\frac{a_2^{J=1}}{b_2^{J=1}} \right)\tsb{CLEO} = -2.27^{+0.57}_{-0.99}
&\stackrel{?}{=}&  \left(\frac{a_2^{J=1}}{b_2^{J=1}} \right)\tsb{th} = 
-2.27 \pm 0.16~, \\
\left(\frac{b_2^{J=2}}{b_2^{J=1}} \right)\tsb{CLEO} =~~ 0.37^{+0.53}_{-0.47}
&\stackrel{?}{=}&  \left(\frac{b_2^{J=2}}{b_2^{J=1}} \right)\tsb{th} = 
1.000 \pm 0.015~, \\
\left(\frac{b_2^{J=2}}{a_2^{J=2}} \right)\tsb{CLEO} = -0.11^{+0.14}_{-0.15}
&\stackrel{?}{=}&  \left(\frac{b_2^{J=2}}{a_2^{J=2}} \right)\tsb{th} = 
-0.297 \pm 0.025~.
\eea

\subsection{$\kappa_c$ calculation}
Our most sensitive measurement of a magnetic quadrupole amplitude is that
of $a_2^{J=1}$.  From theory, we know that
\beq \nonumber
 a_2^{J=1} = -\frac{E_\gamma}{4m_c} (1 + \kappa_c) = (1+\kappa_c)/\xi~,
\eeq
where we defined $1/\xi$ to be the proportionality between $1 + \kappa_c$ and
$a_2^{J=1}$.  If we use $m_c = (1.5 \pm 0.3)$ GeV, we find
$\xi \equiv - (4 m_c)/E_\gamma = -14.0 \pm 2.8$, so
\beq
 1 + \kappa_c = \xi a_2^{J=1} = 0.877 \pm 0.088 \pm 0.034 \pm 0.175~,
\eeq
where we list the result, the statistical uncertainty, the systematic
uncertainty, and the theoretical uncertainty from $m_c = 1.5 \pm 0.3$
GeV.

\subsection{Summary}
We measure significant non-zero magnetic quadrupole amplitudes for the
transitions $\chi_{c1} \to \gamma \jpsi$, $\chi_{c2} \to \gamma \jpsi$, and
$\psip \to \gamp \chi_{c1}$.  Our fits to these three amplitudes all agree well
with the theoretical predictions to first order in the ratio of photon energy
to charmed quark mass with $\kappa_c = 0$ and $m_c = 1.5 \gev$.  The data are
consistent with the lattice QCD prediction (\ref{eqn:chic1lat}) for $\chi_{c1}
\to \gamma \jpsi$), but not with that
for $\chi_{c2} \to \gamma \jpsi$ \cite{Dudek:2009}.  For the transition
$\psip \to \gamp \chi_{c2}$, we do not measure a significant $M2$ amplitude,
though this case has the largest uncertainty since there are fewer $J_{\chi} =
2$ signal events and $E_{\gamp} < E_{\gamu}$ so $|b_2| < |a_2|$.  The non-zero
$M2$ amplitude in the transitions $\chi_{(c1,c2)} \to \gamu \jpsi$ is evident
when comparing the $\cos \theta$ histograms for the data with the histograms
for phase-space-Monte-Carlo events selected to have a pure $E1$ distribution and 
the fitted values of the multipole amplitudes (as shown in Fig.\
\ref{fig:half_scale_costheta_proj}).  We find that for the $J_\chi = 1$ and
$J_\chi =2$ transitions our fitted results differ
from the pure $E1$ value by more than 11$\sigma$ and $6 \sigma$, respectively.  

The agreement between data and theory is in stark contrast in some cases to
previous measurements.  With about 20 times the largest previous data sample,
and a more sophisticated detector, the matter now seems to be resolved.

\section*{ACKNOWLEDGMENTS}
We gratefully acknowledge the effort of the CESR staff in providing us with
excellent luminosity and running conditions.  D.~Cronin-Hennessy 
thanks the A.P.~Sloan Foundation.  J. Rosner thanks the Aspen Center for Physics
for hospitality.  This work was supported by the National Science Foundation,
the U.S. Department of Energy, the Natural Sciences and Engineering Research
Council of Canada, and the U.K. Science and Technology Facilities Council.

\bibliography{prd}
\end{document}